\documentclass[10pt, nolongbibliography,groupedaddress,showkeys,eqsecnum,aps,prd,nofootinbib]{revtex4-2}
\usepackage[colorlinks=true, allcolors=Blue]{hyperref}

\usepackage{amsmath, amssymb, amsfonts}
\usepackage{tensor, mathtools, bm} 
\usepackage{orcidlink}

\usepackage[varbb]{newtx}
\usepackage[dvipsnames]{xcolor}             
\usepackage[]{graphicx} 

\usepackage{microtype}

\begin{document}                                   

\title{Covariant quantization of the Einstein--Hilbert theory in first-order form}

\author{S. Martins-Filho\orcidlink{0000-0003-4195-2713}}   
\email{s.martins-filho@unesp.br}
\affiliation{Instituto de F\'{\i}sica Te\'orica, Universidade Estadual Paulista (UNESP), Rua Dr. Bento Teobaldo Ferraz 271 - Bloco II, 01140-070 S\~ao Paulo, S\~ao Paulo, Brazil}
\date{\today}
\begin{abstract}
We present a covariant quantization of the first-order formulation of the Einstein--Hilbert theory using the path integral and BV formalisms. In this approach, the metric $g^{\mu\nu}$ and the connection $\Gamma^\lambda_{\mu\nu}$ are treated as independent, with the connection playing the role of an auxiliary field. We show that the gauge algebra is closed and irreducible. 
We further demonstrate that the Dyson--Schwinger equations in the first-order formulation lead to structural identities that constrain the Green's functions of the auxiliary field and encode the classical equations of motion at the quantum level. 
We revisit the quantum equivalence between the first- and second-order formulations of the Einstein--Hilbert theory. By employing a suitable trick, a manifestly covariant form of the Senjanović measure is derived. We also show that the two formulations are equivalent at the level of the effective action when the auxiliary field is on-shell.
\end{abstract}
\maketitle

\section{Introduction}

General relativity can be formulated as a field theory of the metric $g^{\mu\nu}$. In the standard formulation, the Einstein--Hilbert (EH) action is given by
\begin{equation}\label{eq:lag}
    S^{(2)}[g] = -\frac{1}{\kappa^2} \int d^4 x \, \sqrt{-g} \, g^{\mu\nu} R_{\mu\nu}(\Gamma),
\end{equation}
where $\kappa^2 = 16\pi G_N$ with $G_N$ the Newton constant, $g = \det g_{\mu\nu}$, and the Ricci tensor is
\begin{equation}
    R_{\mu\nu}(\Gamma) = \partial_{\nu} \Gamma_{\mu\lambda}{}^\lambda{} - \partial_{\lambda} \Gamma_{\mu\nu}{}^\lambda - \Gamma_{\mu\nu}{}^\lambda \Gamma_{\lambda\sigma}{}^\sigma + \Gamma_{\mu\lambda}{}^\sigma \Gamma_{\nu\sigma}{}^\lambda,
\end{equation}
with the Levi-Civita connection
\begin{equation} \label{eq:lcc}
    \Gamma_{\mu\nu}{}^\lambda = \frac{1}{2} g^{\lambda\rho} \left( \partial_{\nu} g_{\mu\rho} + \partial_{\nu} g_{\nu\rho} - \partial_{\rho} g_{\mu\nu} \right).
\end{equation}
From Eq.~\eqref{eq:lag}, one can derive the Einstein field equations, and the theory can be quantized using standard methods for gauge theories.

Note that the Lagrangian \eqref{eq:lag} contains second derivatives of the metric (arising from the Ricci scalar $R = R_{\mu\nu} g^{\mu\nu}$); we therefore refer to this as the \emph{second-order} formulation of the Einstein---Hilbert theory.
However, if the metric $g^{\mu\nu}$ and the connection $\Gamma_{\mu\nu}{}^\lambda$ are treated as independent fields, the same action depends only on first-order derivatives. This defines the \emph{first-order} (Palatini) formulation of the EH theory \cite{Ferraris:1982wci}. In this form, the Feynman rules simplify considerably, yielding only a cubic, momentum-independent interaction vertex. This makes the computation of quantum corrections more manageable \cite{Okubo:1979gt, buchbinder:1983a, buchbinder:1985, McKeon:1994ds, Brandt:2015nxa, Brandt:2016eaj}, and allows for computations of correlation functions of composite fields \cite{McKeon:2020lqp, Brandt:2020vre, Alvarez:2025hym}.

The second-order EH theory can be quantized using the BRST formalism \cite{Becchi:1974xu, Tyutin:1975qk} or the path integral with the Faddeev--Popov procedure \cite{Faddeev:1967fc}, both of which are manifestly covariant. A more general method is the Batalin-Vilkovisky (BV) formalism \cite{Batalin:1981jr, Batalin:1983ggl}, which constructs the quantum action from the gauge algebra and can handle both closed and open algebras, as well as reducible ones \cite{Batalin:1983wj}. In the second-order EH theory, the gauge algebra corresponds to diffeomorphisms, which is closed and irreducible \cite{Buchbinder:2021wzv}. In this case, the BV method reduces to the Faddeev--Popov quantization. 

The covariant quantization of the EH theory in the second-order formulation within the BV formalism has been extensively studied in the literature (see, for example, the review \cite{Gomis:1994he} and more recent works \cite{Barnich:1994db, Bashkirov:2005ig, Brunetti:2013maa, Upadhyay:2013xaa, Cattaneo:2015xca}). 
The BV formalism has also been extended to a variety of more general gravitational theories (see, e.g., Refs.~\cite{Pottel:2020iuz, Kupka:2025hln}), including higher-derivative models\footnote{Higher-derivative gravitational models are known to suffer from ghosts. Interestingly, it has been proposed that such ghost may form bound states, thereby rendering the theory effectively stable~\cite{Asorey:2025bel}.} such as $R^2$ gravity \cite{Stelle:1976gc} and general $f(R)$ theories \cite{Buchdahl:1970ldb, Nojiri:2017ncd, Hu:2023juh, Oikonomou:2025htz}, as well as supergravity \cite{Kallosh:1978de, Nielsen:1978mp}.

Moreover, in the tetrad (first-order) formulation of gravity \cite{Ashtekar:1986yd, BarberoG:1994eia, Holst:1995pc, Immirzi:1996di}, covariant quantization of Palatini gravity has also been received considerable attention \cite{Lavrov:1997xk, Piguet:2000fy, Cattaneo:2017kkv, Brandt:2024rsy}, including recent studies of its renormalization within the BV formalism and the background field method \cite{Brandt:2024kvs}.
In contrast, in the purely metric formulation, to the best of our knowledge, a complete covariant BV treatment has not yet been developed for the EH theory in first-order form.

Motivated by this, in the present work we study the BV quantization of the EH theory in first-order form using the metric as the fundamental field. We adopt the Goldberg parametrization \cite{Goldberg:1958zz},
\begin{equation}
    h^{\mu\nu} = \sqrt{\det g_{\mu\nu}}  g^{\mu\nu},
\end{equation}
and first analyze the gauge algebra of the theory, constructing the corresponding BV quantum action. Quantization is then performed within the path-integral formalism.
The first-order formulation introduces the connection as an auxiliary field, which generates constraints in the Dirac-Bergmann classification \cite{Dirac:1950pj, Bergmann:1949zz}. These are second-class constraints \cite{Kiriushcheva:2006gp, McKeon:2010nf}. While first-class constraints correspond to gauge invariances, second-class constraints impose algebraic restrictions on phase space. Their presence requires the Senjanović procedure \cite{Senjanovic:1976br} rather than the usual Faddeev--Popov approach. Although this has been considered in Ref.~\cite{Chishtie:2012sq}, it leads to a non-manifestly covariant path integral because the canonical structure of the classical theory is itself noncovariant. 

One way to address this issue is the Batalin-Tyutin formalism \cite{Batalin:1991jm}, which converts second-class constraints into first-class ones. Similarly, Faddeev and Shatashvili proposed introducing auxiliary fields, analogous to the Stueckelberg mechanism \cite{stueckelberg1957} for the Proca action, to obtain a covariant measure. 
Another method to obtain a manifestly covariant expression for the Senjanović measure was previously proposed in Ref.~\cite{Filho:2024kxo}. Here, we provide a complete and covariant formulation of this result and extend it by incorporating a generating functional (with sources) for both the graviton and the auxiliary field. 

Moreover, following Ref.~\cite{Lavrov:2021pqh}, we show that the first-order EH theory possesses a novel local symmetry associated with the auxiliary field, which does not alter the quantization due to its triviality.  
We also show that the Dyson--Schwinger equations \cite{Dyson:1949ha, Schwinger:1951ex} leads to structural identities relating the Green's functions of the auxiliary field to its classical value obtained from the equations of motion. Similar identities have been derived and verified at one-loop order in Refs.~\cite{McKeon:2020lqp, Brandt:2020vre} as a consequence of the equivalence between the generating functionals of the second- and first-order formulations. 
The structural identities derived in Ref.~\cite{Brandt:2020vre} are obtained through the relation between the first- and second-order formulations, whereas the identities obtained here follow directly and independently from the Dyson--Schwinger equations of the first-order formulation. As such, they provide a natural quantum realization of the classical equations of motion for the auxiliary field.

Finally, we revisit the quantum equivalence between the first- and second-order formulations of the EH theory. For reviews and detailed discussions, we refer the reader, for instance, to Refs. \cite{Aros:2003bi, Brandt:2020vre, Filho:2024kxo, Brandt:2025lkd}. We show that the Senjanović measure derived here plays an important role in establishing the exact equivalence. We also extend this result to the quantum effective action $\Gamma$ (recently demonstrated for the background effective action in Ref. \cite{Brandt:2025lkd}).

The paper is organized as follows. In Sec.~II, we analyze the gauge algebra of both the first- and second-order formulations, showing that both are closed and irreducible, and that a novel trivial symmetry emerges in the first-order formulation. In Sec.~III, we construct the BV quantum action based on these results. In Sec.~IV, we quantize the first-order theory, first starting from the generating functional of the second-order formulation and then directly in first-order form, discussing their quantum equivalence. We conclude this section with a discussion of the BRST symmetry of the EH theory. 
In Sec.~V, we derive the Dyson--Schwinger identities from the general invariance of the generating functional and show that they generate intrinsic structural identities. Finally, in Sec.~VI, we present a brief discussion of our results. In Appendix~A, we analyze the contributions of the Senjanović measure at finite temperature. The Zinn-Justin master equations~\cite{Zinn-Justin:1974ggz} for the EH theory in both first- and second-order formulations are derived in Appendix~B. Throughout this paper, right and left functional derivatives with respect to a field (or antifield) are denoted by ``$\leftarrow$'' and ``$\rightarrow$'', respectively.

\section{Gauge algebra}
In this section, we analyze the gauge algebra of both the first and second order formulations of the EH theory. We also show that these theories are equivalent at the classical level. It is important to note that the quantization in the BV formalism depends on the structure of the gauge algebra. When the algebra is closed and irreducible, the BV formalism reduces to the Faddeev--Popov quantization.

\subsection{Second-order formulation}
We use the Goldberg metric $h^{\mu \nu} \equiv  \sqrt{-g} g^{\mu \nu}$. We will also consider a weak field expansion $  h^{\mu \nu} = \eta^{\mu \nu} + \kappa \mathfrak{h}^{\mu \nu} $. The standard formulation of the EH theory using this parametrization can be written as [see Eq. (3.13) of Ref.~\cite{Brandt:2016eaj}]
\begin{equation}\label{eq:EH2}
S^{(2)}[h] 
= - \frac{1}{2\kappa^{2}} \int \mathop{d x} \partial_{\lambda} \tensor{h}{^{\mu \nu} }
    \tensor{(M^{-1})}{_{\mu \nu}^{\lambda}_{\alpha \beta}^{\gamma} }(h) 
    \partial_{\gamma} \tensor{h}{^{\alpha \beta}}, 
\end{equation}
where, in ${D}$ spacetime dimensions,
\begin{equation}\label{eq:definvM}
\begin{split}
    \tensor{(M^{-1})}{_{\mu \nu}^{\lambda}_{ \pi \tau}^{ \rho}} (h) 
={}&- \frac{1}{2({{D}}-2)} h^{\lambda \rho} h_{\mu \nu} h_{\pi \tau} +
    \frac{1}{4} h^{\lambda \rho} \left ( h_{\pi \mu} h_{\tau \nu} + h_{\pi \nu} h_{\tau \mu}\right ) 
    \\
    & - \frac{1}{4} \left ( h_{\tau \mu} \delta_{\nu}^{\rho} \delta_{\pi}^{\lambda} + h_{\pi \mu} \delta_{\nu}^{\rho} \delta_{\tau}^{\lambda} +  h_{\tau \nu} \delta_{\mu}^{\rho} \delta_{\pi}^{\lambda} +  h_{\pi \nu} \delta_{\mu}^{\rho} \delta_{\tau}^{\lambda}\right ).
\end{split}
\end{equation}
The Einstein equations of motion can be derived from Eq.~\eqref{eq:EH2}, which reads 
\begin{equation}\label{eq:EOM2}
    \frac{1}{2} \partial_{\lambda} \tensor{h}{^{\mu \nu}} 
    \tensor{(M^{-1})}{_{\mu \nu}^{\lambda}_{\alpha' \beta'}^{\gamma'} }(h) 
    \frac{\partial \tensor{M}{^{\alpha ' \beta '}_{\gamma '}^{\mu ' \nu '}_{\lambda '}} (h)}{\partial h^{\pi \tau}} 
    \tensor{(M^{-1})}{_{\mu' \nu'}^{\lambda'}_{\alpha \beta}^{\gamma} }(h) 
    \partial_{\gamma}  \tensor{h}{^{\alpha \beta}}
-\partial_{\lambda } [\tensor{(M^{-1})}{_{\pi \tau }^{\lambda}_{\alpha \beta}^{\gamma} }(h) 
\partial_{\gamma} \tensor{h}{^{\alpha \beta}}]=0,
\end{equation}
where 
\begin{equation}\label{eq:defM}
    \begin{split}
        \tensor{M}{^{\mu\nu}_{\lambda}^{\pi\tau}_{\sigma}}(h)   &= 
        \frac{1}{2}\Big[\frac{1}{D-1}\left( \delta^\nu_\lambda\delta^\tau_\sigma h^{\mu\pi}+
                                                \delta^\mu_\lambda\delta^\tau_\sigma h^{\nu\pi}+
                                                \delta^\nu_\lambda\delta^\pi_\sigma h^{\mu\tau}+
                                                \delta^\mu_\lambda\delta^\pi_\sigma h^{\nu\tau}
\right) 
      \\  & \qquad \quad  
      -  
\left( 
                                                \delta^\tau_\lambda\delta^\nu_\sigma h^{\mu\pi}+
                                                \delta^\tau_\lambda\delta^\mu_\sigma h^{\nu\pi}+
                                                \delta^\pi_\lambda\delta^\nu_\sigma h^{\mu\tau}+
                                                \delta^\pi_\lambda\delta^\mu_\sigma h^{\nu\tau}
                                        \right) \Big]
\end{split}
\end{equation}
is the inverse of $ M^{-1}$.

The action \eqref{eq:EH2} is invariant under the gauge transformations (diffeomorphisms)
\begin{equation}\label{eq:gt2}
\begin{split}
    \delta_{\zeta} h^{\mu \nu}(y) ={}& {h}^{\lambda \nu} \partial_{\lambda} \zeta^{\mu} + {h}^{\mu \lambda} \partial_{\lambda} \zeta^{\nu} - \partial_{\lambda} ( {h}^{\mu \nu} \zeta^{\lambda} ) \\
    \equiv{}& \int \mathop{d x} \tensor{\mathcal{R}}{^{\mu \nu}_{\lambda}} (x,y;h)\zeta^{\lambda}(x); 
\end{split}
\end{equation}
where  $ \zeta^{\mu} (x)$ is a infinitesimal gauge parameter and 
\begin{equation}\label{eq:defR}
    \tensor{\mathcal{R}}{^{\mu \nu}_{\lambda}} (x,y;h)=
    \left({h}^{\sigma \nu} \partial_{\sigma} \delta_{\lambda}^{\mu}  + {h}^{\mu \sigma} \partial_{\sigma} \delta_{\lambda}^{\nu}   - \partial_{\sigma}  {h}^{\mu \nu} \delta_{\lambda}^{\sigma} \right)\delta (x-y).
\end{equation}

The gauge algebra of the EH theory is given by
\begin{equation}\label{eq:ga2s}
[\delta_{\zeta_1}, \delta_{\zeta_2}] h^{\mu \nu} (x) = \int \mathop{d w} \tensor{\mathcal{R}}{^{\mu \nu}_{\lambda}} (x,w;h) \zeta^\lambda_3(w),
\end{equation}
where
\begin{equation} \label{eq:zeta3}
\zeta^\lambda_3(w)= \int \mathop{d y} \mathop{d z} \tensor{f}{^{\lambda}_{\alpha}_{\beta}}(w,y,z) {\zeta}^\alpha_1(y){\zeta}^\beta_2(z)
\end{equation}
and
\begin{equation} \label{eq:deff}
    \tensor{f}{^{\lambda}_{\alpha \beta}}     (x,y,z)= \delta (x-y) \delta_{\beta}^{\lambda} \partial_{\alpha} \delta (x-z) - \delta (x-z) \delta_{\alpha}^{\lambda} \partial_{\beta} \delta (x-y).
\end{equation}
It is well-known that this algebra is closed and irreducible \cite{Buchbinder:2021wzv}.

\subsection{First-order formulation}
The first order formulation of the EH gravity reads \cite{McKeon:2010nf} 
\begin{equation}\label{eq:EH1}
S^{(1)}[h, G] = \frac{1}{\kappa^{2}} \int \mathop{d x} \left ( 
    \frac{1}{2}\tensor{G}{_{\mu \nu}^{\lambda}} \tensor{M}{^{\mu \nu}_{\lambda}^{\alpha \beta}_{\gamma}} \tensor{G}{_{\alpha \beta}^{\gamma}} 
    -  \tensor{G}{_{\mu \nu}^{\lambda}} \partial_{\lambda} h^{\mu \nu}  
\right )
\end{equation}
where $ \tensor{G}{_{\mu \nu}^{\lambda}} $ is an auxiliary field and $ h^{\mu \nu} = \eta^{\mu \nu} + \kappa \mathfrak{h}^{\mu \nu} $. 
The equations of motion read
\begin{subequations}\label{eq:EOM}
\begin{equation}\label{eq:EOM0}
\frac{1}{2}\tensor{G}{_{\mu \nu}^{\lambda}} \frac{\partial \tensor{M}{^{\mu \nu}_{\lambda}^{\alpha \beta}_{\gamma}}}{\partial h^{\pi \tau} }\tensor{G}{_{\alpha \beta}^{\gamma}} 
+ \partial_{\lambda}\tensor{G}{_{\pi \tau }^{\lambda}} =0,
\end{equation}
\begin{equation}\label{eq:EOM1}
\tensor{M}{^{\mu \nu}_{\lambda}^{\alpha \beta}_{\gamma}} \tensor{G}{_{\alpha \beta}^{\gamma}} -   \partial_{\lambda} h^{\mu \nu} =0. 
\end{equation}
\end{subequations}
One can use Eq.~\eqref{eq:lcc} to show that the Eq.~\eqref{eq:EOM1} is equivalent to the classical value of $ \tensor{G}{_{\mu \nu}^{\lambda}} $, which is given by
\begin{equation}\label{eq:valueG}
    \mathcal{G}_{\mu \nu}{}^{\lambda} [h]\equiv \tensor{\Gamma}{_{\mu \nu}^{\lambda}} -     
    \frac{1}{2} \left( \delta_{\mu}^{\lambda} \Gamma{}_{\nu \alpha }{}^{\alpha} + \delta_{\nu}^{\lambda } \Gamma{}_{\mu \alpha } {}^{\alpha}\right).
\end{equation}
Substituting Eq.~\eqref{eq:valueG} in the first-order Lagrangian \eqref{eq:EH1}, we obtain the EH Lagrangian in second order form~\eqref{eq:EH2}. Note that, using the second equation of Eq.~\eqref{eq:valueG} into the first equation leads to the equations of motion of the second-order formulation shown in Eq.~\eqref{eq:EOM2}. This shows the classical equivalence between the first and second-order formulations of the EH theory. 

The action \eqref{eq:EH1} is invariant under the gauge transformations \eqref{eq:gt2},
\begin{equation}\label{eq:gt1}
\delta_{\zeta}S^{(1)}[h, G] =0,
\end{equation}
as long as the auxiliary field transforms as its classical value: 
\begin{equation}\label{eq:transG}
        \delta_{\zeta} \tensor{G}{_{\mu \nu}^{\lambda}}(x) = \int \mathop{d w} \tensor{\mathcal{R}}{_{\mu \nu}^{\lambda}_{\rho}} (x,w;G) \zeta^{\rho} (w) 
    ,
\end{equation}
where 
\begin{equation}\label{eq:defRG}
    \begin{split}
        \tensor{\mathcal{R}}{_{\mu \nu}^{\lambda}_{\rho}} (x,w;G) \equiv
        \kappa \bigg[ & - \partial_{\mu} \partial_{\nu} \delta^{\lambda}_{\alpha}  + \frac{1}{2} ( \delta_{\mu}^{\lambda} \partial_{\nu} + \delta_{\nu}^{\lambda} \partial_{\mu} ) \partial_{\rho} \delta^{\rho}_{\alpha}   - \delta^{\rho}_{\alpha} \partial_{\rho}  \tensor{G}{_{\mu \nu}^{\lambda}}    + \tensor{G}{_{\mu \nu}^{\rho}} \partial_{\rho} \delta^{\lambda}_{\alpha}  - ( \tensor{G}{_{\mu \rho}^{\lambda}} \partial_{\nu} + \tensor{G}{_{\nu \rho}^{\lambda}} \partial_{\mu} )\delta^{\rho}_{\alpha} \bigg] \delta (x-w).
\end{split}
\end{equation}

Additionally, it is also invariant under the novel local symmetry:
\begin{equation}\label{eq:gt1ad}
\delta_{ \xi} h^{\mu \nu} =0,\quad
\delta_{ \xi} \tensor{G}{_{\mu \nu}^{\alpha}}(z) = \int \mathop{d x} \mathop{d y}  \tensor{f}{_{\lambda}^{\alpha \beta}}(x,y,z) [G - \mathcal{G} ] \tensor{}{_{\mu \nu}_{\, \beta}} (y) \xi^{\lambda}(x).
\end{equation}
where ${ \xi}^\lambda $ is an arbitrary gauge parameter. 
Note that the transformation of the auxiliary field can be rewritten as 
\begin{equation}\label{eq:defG}
    \delta_{ \xi} \tensor{G}{_{\mu \nu}^{\beta}}(z) = \int \mathop{d x} \tensor{f}{_{\lambda}^{\alpha \beta}}(x,y,z) \tensor{(M^{-1})}{_{\mu \nu}_{\alpha \, }_{\mu ' \nu '}^{\alpha '}} \frac{\delta S^{(1)}}{\tensor{G}{_{\mu' \nu'}^{\alpha'}}}  \xi^{\lambda}(x).
\end{equation}
This shows that the auxiliary field transforms proportionally to its equation of motion. Thus, the invariance of the action under this transformation is trivial \cite{Henneaux:1989jq}, as shown below: 
\begin{equation}\label{eq:invS}
    \begin{split}
        \delta_{\xi} S^{(1)} [h,G] ={}& \int \mathop{d z} \frac{\delta S^{(1)}}{\delta \tensor{G}{_{\mu \nu}^{\beta}}(z)} \delta_{\xi} \tensor{G}{_{\mu \nu}^{\beta}}(z)  \\ ={}& \int \mathop{d x} \mathop{d y} \mathop{d z}  \tensor{f}{_{\lambda}^{\alpha \beta}} (x,y,z) \tensor{(M^{-1})}{_{\mu \nu}_{\alpha \,  }_{\mu ' \nu '}^{\alpha '}} \tensor{(M^{-1})}{^{\mu \nu }_{\beta \, }_{\mu ' \nu '}^{\beta '}} \frac{\delta S^{(1)}}{\delta G_{\mu '\nu '}{}^{\alpha'}(z)} 
        \frac{\delta S^{(1)}}{\delta G_{\mu '\nu '}{}^{\beta '}(x)} \xi^{\lambda} (x) =0,
    \end{split} 
\end{equation}
where we have used that $ \tensor{f}{_{\lambda}^{\alpha \beta}} = - \tensor{f}{_{\lambda}^{\beta \alpha}} $.
This shows that the novel local symmetry \eqref{eq:gt1ad} is not a genuine gauge symmetry, i.e., it is not associated with any degeneracy of the EH action.

The gauge algebra of the first order formulation of EH follows
\begin{align}
\label{gaugeal1}
[\delta_{\xi_1}, \delta_{\xi_2}] h^{\mu \nu} (x)={}& 
\int \mathop{d w} \tensor{\mathcal{R}}{^{\mu \nu}_{\lambda}} (x,w;h)\zeta_{3}^{\lambda} (w),\\
[\delta_{\xi_1}, \delta_{\xi_2}] \tensor{G}{_{\mu \nu}^{\lambda}} (x)={}&
\int \mathop{d w} \tensor{\mathcal{R}}{_{\mu \nu}^{\lambda}_{\alpha}}(x,w;G) \zeta_{3}^{\alpha} (w).
\end{align}
We remind that $ \zeta_{3} $ is the same parameter that appears in the second order gauge algebra and it is defined in Eq.~\eqref{eq:zeta3}. Similarly, as in the second order case, the first order gauge algebra \eqref{gaugeal1} is closed and irreducible.

\section{BV formalism}

In order to derive the quantum action for EH theory in the first order form using the BV-formalism \cite{Batalin:1981jr, Batalin:1983ggl}, we introduce the set of fields ${\bm{\phi}}^I_{}$ and antifields $\prescript{\star}{}{\bm{\phi}}^{I}_{}$ (which forms an antisymplectic space)
\begin{equation}
{\bm{\phi}}^I_{}=\begin{pmatrix}  {}\mathfrak{h}^{\mu \nu} &  \tensor{G}{_{\mu \nu}^{\lambda}} & d^{\mu}  \end{pmatrix},\quad
\prescript{\star}{} {\bm{\phi}}^I_{}= \begin{pmatrix} 
    \prescript{\star}{}  {}\mathfrak{h}^{ \mu \nu} & \prescript{\star}{} G\tensor{}{_{\mu \nu}^{\lambda}} & \prescript{\star}{} d^{\mu} 
\end{pmatrix},
\end{equation}
where $d^{\mu} $  is the Faddeev--Popov ghost field. The antifields are the sources of the BRST transformations of the corresponding fields.

Now, we have to find a solution to the master equation
\begin{equation}
(S_{},S_{}) \equiv 
\int \mathop{d^{4} x} S\frac{\overset{\leftarrow}{\delta}  }{\delta {\bm{\phi}}^{I}  } \frac{\overset{\rightarrow}{\delta}  }{\delta \prescript{\star}{} {} {\bm{\phi}}^{I} }S =0
\end{equation}
that satisfies the boundary condition
\begin{equation}
S_{}\big|_{ \prescript{\star}{} {\bm{\phi}}_{}=0}=S^{(1)}[h,G].
\end{equation}
We recall that the gauge algebra \eqref{eq:gt1} is closed and irreducible. Thus, the quantum action is linear in the antifields, which reads
\begin{equation}
    \begin{split}
        S   ={}&S^{(1)}[h,G]
+ \prescript{\star}{} h_{\mu \nu} [ {}\mathfrak{h}^{\sigma \nu} \partial_{\sigma} d^{\mu} +  {}\mathfrak{h}^{\mu \lambda} \partial_{\lambda} d^{\nu} - \partial_{\lambda} (  {}\mathfrak{h}^{\mu \nu} d^{\lambda } )] 
        - \prescript{\star}{} d_{\mu} d^{\lambda} \partial_{\lambda} d^{\mu}  
                       \\ & + \kappa \prescript{\star}{} G \tensor{}{^{\mu \nu}_{\lambda}} 
        \bigg[  - \partial_{\mu} \partial_{\nu} \delta^{\lambda}_{\alpha}  + \frac{1}{2} ( \delta_{\mu}^{\lambda} \partial_{\nu} + \delta_{\nu}^{\lambda} \partial_{\mu} ) \partial_{\rho} \delta^{\rho}_{\alpha}   - \delta^{\rho}_{\alpha} \partial_{\rho}  \tensor{G}{_{\mu \nu}^{\lambda}}    + \tensor{G}{_{\mu \nu}^{\rho}} \partial_{\rho} \delta^{\lambda}_{\alpha}  - ( \tensor{G}{_{\mu \rho}^{\lambda}} \partial_{\nu} + \tensor{G}{_{\nu \rho}^{\lambda}} \partial_{\mu} )\delta^{\rho}_{\alpha} \bigg]  d^{\lambda} 
        .
    \end{split}
\end{equation}

Next, we have to introduce a gauge fixing term. For this, we extend the action by introducing the Nakanishi-Lautrup field $ B^{\mu} $ \cite{lautrup:1967, Nakanishi:1966zz}, the FP antighost $ \bar{d}^{\mu} $ and their corresponding antifields. 
The resulting action $S'=S[{\bm{\phi}}^{(1)}, \prescript{\star}{} {\bm{\phi}}^{(1)}]$ in the extended antisymplectic space 
\begin{equation}
{\bm{\phi}}^{(1) \, I}=
\begin{pmatrix}
     {}\mathfrak{h}^{\mu \nu} & \tensor{G}{_{\mu \nu}^{\lambda}} & d^{\mu} & \bar{d}^{\mu} & B^{\mu} 
\end{pmatrix}
,\quad
\prescript{\star}{} {} {\bm{\phi}}^{(1) \, I}=
\begin{pmatrix}
    \prescript{\star}{} {}  {}\mathfrak{h}^{\mu \nu}  & \prescript{\star}{} {} \tensor{G}{_{\mu \nu}^{\lambda}} & \prescript{\star}{} {} d^{\mu} & \prescript{\star}{} {} \bar{d}^{\mu} & \prescript{\star}{} {} B^{\mu} 
\end{pmatrix}
\end{equation}
is given by
\begin{equation} \label{eq:fullS}
    \begin{split}
        S'  =S + \prescript{\star}{} {} \bar{d}^{\mu} B_{\mu}.
\end{split}
\end{equation}
The action \eqref{eq:fullS} also satisfies the master equation 
\begin{equation}
(S',S')=0, \quad S'\big|_{\prescript{\star}{} {} {\bm{\phi}}^{(1)}=0}=S^{(1)}[h,G].
\end{equation}

We obtain the gauge fixed action $S_{\text{FP}}[{\bm{\phi}}]$ by introducing a BRST exact fermionic function as follows:
\begin{equation}
    S_{\text{FP}}^{(1)}[{\bm{\phi}}^{(1)}]=S'\left[{\bm{\phi}}^{(1)},\prescript{\star}{} {} {\bm{\phi}}^{(1)}=
\Psi\overset{\leftarrow}{\partial}_{{\bm{\phi}}^{(1)}}\right],
\end{equation}
where $\Psi=\Psi[{\bm{\phi}}^{(1)}]$ is the gauge-fixing potential. 
Here, we will consider the De Donder gauge for the Goldberg metric. Thus, the corresponding gauge-fixing potential in a general gauge reads
\begin{equation} \label{eq:newl}
\Psi={\bar d}_\mu
\left(\partial_{\nu}  {}\mathfrak{h}^{\mu \nu} +\frac{\xi}{2}B^\mu\right),
\end{equation}
where $ \xi $ is a constant (analogous to $R_{\xi} $ gauges in YM theories). When $ \xi =0$, we have the Landau-DeWitt gauge, which is the singular case of Eq.~\eqref{eq:newl}.  In the Landau-DeWitt gauge, the quadratic operator assumes minimal form and the effective action remains manifestly background gauge invariant, greatly simplifying one-loop analyses.

The gauge fixed action for the EH theory in first order form \eqref{eq:EH1} is then given by 
\begin{equation}\label{eq:FP1}
\begin{split}
    S_{\text{FP} }^{(1)} & [{\bm{\phi}}^{(1)}] =S^{(1)}[h,G] 
+ B^\mu\left(\partial_{\nu}  {}\mathfrak{h}^{\mu \nu} +\frac{\xi}{2}B^\mu\right)
                                     \\ & 
                                     + \bar d_\mu\left\{\partial^2 \eta^{\mu\nu} + \kappa \left[( \partial_{\rho} \mathfrak{h}^{\rho\sigma})\partial_\sigma\eta^{\mu\nu}-(\partial_{\rho} \mathfrak{h}^{\rho\mu})\partial^{\nu} +\mathfrak{h}^{\rho\sigma}\partial_{\rho}\partial_\sigma\eta^{\mu\nu}-(\partial_\rho\partial^\nu\mathfrak{h}^{\rho\mu})\right]\right\} d_\nu.
\end{split}
\end{equation}

In the second-order formulation, the quantum action is similar:
\begin{equation}\label{eq:FP2}
\begin{split}
    S_{\text{FP} }^{(2)} & [{\bm{\phi}}^{(2)}] = S^{(2)}[h] 
+ B^\mu\left(\partial_{\nu}  {}\mathfrak{h}^{\mu \nu} +\frac{\xi}{2}B^\mu\right)
                                     \\ & 
                                     + \bar d_\mu\left\{\partial^2 \eta^{\mu\nu} + \kappa \left[( \partial_{\rho} \mathfrak{h}^{\rho\sigma})\partial_\sigma\eta^{\mu\nu}-(\partial_{\rho} \mathfrak{h}^{\rho\mu})\partial^{\nu} +\mathfrak{h}^{\rho\sigma}\partial_{\rho}\partial_\sigma\eta^{\mu\nu}-(\partial_\rho\partial^\nu\mathfrak{h}^{\rho\mu})\right]\right\} d_\nu.
\end{split}
\end{equation}
where
${\bm{\phi}}^{(2) \, I}= \begin{pmatrix}
    \mathfrak{h}^{\mu \nu} & d^{\mu} & \bar{d}^{\mu} & B^{\mu}
\end{pmatrix}
$ are fields of the extended antisymplectic
space in this form. Comparing the actions \eqref{eq:FP1} and \eqref{eq:FP2}, we see that the Faddeev--Popov sector is equal in both cases. This is an interesting characteristic that can be explored to derive the quantum equivalence. However, the first-order form of gauge theories may accommodate more general gauge fixings due to the auxiliary field. Here, we will only consider the case of that $ \Psi $ depends on the graviton field $ \mathfrak{h}^{\mu \nu} $ and $ B^{\mu} $.

\section{Quantization}
Now, we are ready to proceed with the quantization using the path integral formalism.
The generating functional of the first order formulation of the EH theory is given by 
\begin{equation}\label{eq:Z2}
    Z^{(2)}[ \bm{j}  ]= \frac{1}{N} \int \mathcal{D} {\bm{\phi}}^{(2)}
    \exp\left[i\left(S_{\text{FP} }^{(2)}[{\bm{\phi}}^{(2)}] + \int \mathop{d x} \bm{j}  \cdot \bm{\mathfrak{h} }   \right)\right].
\end{equation}
where $N$ is a normalization factor, $ \bm{j} =\{j_{ \mu \nu }\}$ is the external source for the field $ \mathfrak{h}^{\mu \nu} $ in matrix form (similar notation is employed to the fields $ \bm{\mathfrak{h}} = \{ \mathfrak{h}^{\mu \nu} \}$) and $ \bm{A} \cdot \bm{B}  \cdot \bm{C} \equiv  \mathop{\rm Tr} (\bm{ABC})$. 

To obtain the generating functional in the first-order form, we employ a trick proposed in \cite{Filho:2024kxo}. The procedure starts by inserting the constant factor
\begin{equation}\label{eq:1HE}
    1 =
    \int 
    \mathop{\mathcal{D} \bm{G}} 
    \Delta^{1/2}( h,J)  \exp i \int \mathop{dx} \left ( 
        \frac{1}{2 \kappa^{2} } \bm{G} \cdot  
        \bm{M} 
        \cdot 
        \bm{G} 
        + \bm{J} \cdot \bm{G} 
    \right)
\end{equation}
into an extension of the generating functional \eqref{eq:Z2}:
\begin{equation}\label{eq:ZHEwithSourceG}
    Z^{(2)}[ \bm{j} ; \bm{J} ]= \frac{1}{N} \int \mathcal{D} {\bm{\phi}}^{(2)}
    \exp\left\{i\left[S_{\text{FP}}^{(2)}[{\bm{\phi}}^{(2)}] + \int \mathop{d x} \left(\bm{j}  \cdot \bm{\mathfrak{h} } + \bm{J} \cdot \bm{\mathcal{G}}
   -\frac{\kappa^{2} }{2} \bm{J} \cdot \bm{M}^{-1} \cdot \bm{J}     \right)\right]\right\}
\end{equation}
in which an additional source for the classical value of the auxiliary field $ \mathcal{G}^{\lambda}_{ \mu \nu }(h)$ is introduced. 

Then, we redefine the field $ \tensor{G}{_{\mu \nu}^{\lambda}} $ by a shift:
\begin{equation}\label{eq:redfG}
    \bm{G}  \to \bm{G}  
    - \bm{\mathcal{G}} 
\end{equation}
leading to
\begin{equation}\label{eq:fgHE12}
    \begin{split}
        \frac{1}{N} \int   & \mathop{\mathcal{D} {\bm{\phi}}^{(1)} }
                            \Delta^{1/2} (h,J) 
                            \exp\left\{i\left[S_{\text{FP} }^{(1)}[{\bm{\phi}}^{(1)}] + \int \mathop{d x} \left(\bm{j}  \cdot \bm{\mathfrak{h} }  + \bm{J}  \cdot \bm{G}   \right)\right]\right\}.
\end{split} 
\end{equation}
The new factor $ \Delta^{1/2}  (h,J)$ corresponds to the Senjanović determinant \cite{Senjanovic:1976br}. To derive the explicit form of this factor, we integrate the auxiliary field in Eq.~\eqref{eq:1HE} arriving at
\begin{equation}\label{eq:DeltaHE}
    \Delta^{1/2} (h,J) = 
|\det \tensor*{M}{*_{ \lambda}^{\mu \nu}_{\rho}^{\pi \tau}}  (h)|^{1/2} 
\exp i \int \mathop{d x} \left (  
\frac{\kappa^{2}}{2} \bm{J} \cdot \bm{M}^{-1} \cdot \bm{J} \right ),
\end{equation}
where the first term is the Senjanović determinant. This determinant arises due to the presence of second-class constraints associated with the auxiliary field $ \tensor{G}{_{\mu \nu}^{\lambda}} $ \cite{McKeon:2010nf}. Using this trick, we obtain a manifestly covariant result for the Senjanović determinant (cf.\ with the measure in Eq. (64) of Ref. \cite{Chishtie:2012sq}). 

Now, substituting Eq.~\eqref{eq:DeltaHE} into Eq.~\eqref{eq:fgHE12}, we find the generating functional in first order form:
\begin{equation}\label{eq:Z1}
    Z^{(1)}[ \bm{j} , \bm{J} ]= 
    \frac{1}{N} \int \mathcal{D} {\bm{\phi}}^{(1)}
    |\det \bm{M}(h) |^{1/2}
    \exp\left\{i\left[S_{\text{FP} }^{(1)}[{\bm{\phi}}^{(1)}] + \int \mathop{d x} \left(\bm{j}  \cdot \bm{\mathfrak{h} }  + \bm{J}  \cdot \bm{G} \right)\right]\right\},
\end{equation}
where $\bm{j}=\{j^{\mu\nu}\}$ and $\bm{J}=\{\tensor{J}{^{\mu\nu}_{\lambda}}\}$ denote external sources written in matrix notation. An analogous notation is used for the auxiliary field  $\bm{G}=\{\tensor{G}{_{\mu\nu}^{\lambda}}\}$. The resulting expression is manifestly covariant, in contrast with formulations obtained by applying the Senjanovi\'c-Faddeev procedure directly (see, for example, Ref.~\cite{Chishtie:2012sq, Aros:2003bi}).

Note that in the literature, the Senjanović determinant is usually ignored \cite{Chishtie:2011wd} due to its apparent triviality. Indeed, $\det \bm{M}(h)$ gives rise to a set of massless tadpole-like contributions that vanish in dimensional regularization \cite{Brandt:2020gms}. However, as we have demonstrated, it plays an important role in understanding the quantum equivalence of the two formulations. At finite temperature (see Appendix~\ref{sec:finite}), this remains true, since the determinant is required to cancel the contributions from the auxiliary field in the free energy $F$ at functional level. 

\subsection{Quantum equivalence}\label{sec:QE}

Therefore, we find the first order form of the 
generating functional of the second-order formulation EH theory \eqref{eq:Z2}. The quantum equivalence follows as a consequence of the general relation
\begin{equation}\label{eq:qequivgeneral}
    Z^{(1)} [ \bm{j} , \bm{J} ] = Z^{(2)} [ \bm{j} ; \bm{J} ]
\end{equation}
that implies that 
\begin{equation}\label{eq:qequiv}
    Z^{(1)} [ \bm{j} , 0] = Z^{(2)} [ \bm{j}  ].
\end{equation}
This means that the first- and second-order formulations of the EH theory share identical gauge field Green’s functions: 
\begin{equation}\label{eq:equivh}
    \langle 0|T \mathfrak{h}^{\mu_{1} \nu_{1}} (x_{1} ) \cdots \mathfrak{h}^{\mu_{n} \nu_{n}} ( x_{n} )| 0 \rangle^{(1)} =
    \langle 0|T \mathfrak{h}^{\mu_{1} \nu_{1}} (x_{1} ) \cdots \mathfrak{h}^{\mu_{n}  \nu_{n}} ( x_{n} )| 0 \rangle^{(2)},
\end{equation}
where the labels $(1)$ and $(2)$ mark the Green's function computed in the first and second order formulations respectively.

Moreover, it has been shown that the use of a relation analogous to
Eq.~\eqref{eq:qequivgeneral} (see Refs.~\cite{McKeon:2020lqp, Brandt:2020vre}), allows one to derive several structural identities that connect Green’s functions computed in the first-order formulation with their counterparts in the second-order formulation. For example, we have that 
\begin{subequations}\label{eq:SID0}
\begin{align}\label{eq:SID10}
    & \langle 0|T \tensor{G}{_{\mu \nu}^{\lambda}} (x) \tensor{G}{_{\pi \tau}^{\gamma} } (y)| 0 \rangle^{(1)} 
    = i \kappa^{2} \delta (x-y)\langle 0|T \tensor{(M^{-1})}{_{\mu \nu}^{\lambda}_{\pi \tau}^{\gamma}}(x)| 0 \rangle^{(2)} 
    +  \langle 0|T \tensor{\mathcal{G}}{_{\mu \nu}^{\lambda}} (x) \tensor{\mathcal{G}}{_{\pi \tau}^{\gamma} } (y)| 0 \rangle^{(2)} 
    \\ \label{eq:SID20}
    & \langle 0|T \tensor{G}{_{\mu \nu}^{\lambda} } (x) \tensor{\mathfrak{h}}{^{\alpha \beta}} (y)| 0 \rangle^{(1)} 
    =
    \langle 0|T \tensor{\mathcal{G} }{_{\mu \nu}^{\lambda} } (x) \tensor{\mathfrak{h}}{^{\alpha \beta}} (y)| 0 \rangle^{(2)}.
\end{align}
\end{subequations}
These structural identities are complementary to the usual Slavnov--Taylor \cite{Slavnov:1972fg, Taylor:1971ff} identities satisfied by gauge theories. 

Now, one can use the structural identity \eqref{eq:equivh} to derive the following relations: 
\begin{subequations}\label{eq:SID}
\begin{align}\label{eq:SID1}
    & \langle 0|T \tensor{G}{_{\mu \nu}^{\lambda}} (x) \tensor{G}{_{\pi \tau}^{\gamma} } (y)| 0 \rangle^{(1)} 
    = i \kappa^{2} \delta (x-y)\langle 0|T \tensor{(M^{-1})}{_{\mu \nu}^{\lambda}_{\pi \tau}^{\gamma}}(x)| 0 \rangle^{(1)} 
    +  \langle 0|T \tensor{\mathcal{G}}{_{\mu \nu}^{\lambda}} (x) \tensor{\mathcal{G}}{_{\pi \tau}^{\gamma} } (y)| 0 \rangle^{(1)} 
    \\ \label{eq:SID2}
    & \langle 0|T \tensor{G}{_{\mu \nu}^{\lambda} } (x) \tensor{\mathfrak{h}}{^{\alpha \beta}} (y)| 0 \rangle^{(1)} 
    =
    \langle 0|T \tensor{\mathcal{G} }{_{\mu \nu}^{\lambda} } (x) \tensor{\mathfrak{h}}{^{\alpha \beta}} (y)| 0 \rangle^{(1)}.
\end{align}
\end{subequations}
Note that, these relations become a direct relation between the Green's functions (computed in the first-order formulation) of the auxiliary field $ \tensor{G}{_{\mu \nu}^{\lambda}} $ and its classical value $ \tensor{\mathcal{G}}{_{\mu \nu}^{\lambda}} ( \mathfrak{h} )$. 
In the Sect. V, we show that these stronger identities (and many others) can be derived from the Dyson--Schwinger equation for the auxiliary field.

By employing the expression
\begin{equation}\label{eq:detSexp}
    \det \bm{M}^{1/2} = \int 
    \mathop{\mathcal{D} \bm{H} } \mathop{\mathcal{D} \bm{\theta} }  \mathop{\mathcal{D} \bar{\bm{{\theta}}}} \exp \int \mathop{d x} {\left ( \frac{1}{2} \bm{H} \cdot \bm{M} \cdot \bm{H} + \bar{\bm{\theta}} \cdot \bm{M}  \cdot \bm{\theta} \right )}
\end{equation}
to exponentiate the Senjanović determinant in Eq.~\eqref{eq:Z1}, we obtain the local form of the generating functional of the EH theory in first order: 
\begin{equation}\label{eq:Z1local}
     Z^{(1)}[ \bm{j} , \bm{J} ]=\int \mathcal{D} {\bm{\phi}}^{(1)}
 \mathop{\mathcal{D} \bm{H} } \mathop{\mathcal{D} \bm{\theta} }  \mathop{\mathcal{D} \bar{\bm{{\theta}}}}
 \exp\left\{i\left[S_{\text{FP} }^{(1)}[{\bm{\phi}}^{(1)}] + \int \mathop{d x} \left(\frac{1}{2} \bm{H} \cdot \bm{M}(h) \cdot \bm{H} + \bar{\bm{\theta}} \cdot \bm{M} (h) \cdot \bm{\theta} +   
    \bm{j}  \cdot \bm{\mathfrak{h} }  + \bm{J}  \cdot \bm{G} \right)\right]\right\},
\end{equation}
where $ \tensor{H}{_{\mu \nu}^{\lambda}} $ is a bosonic field and $ \tensor{\theta}{_{\mu \nu}^{\lambda}} $, $ \tensor{\bar{\theta}}{_{\mu \nu}^{\lambda}} $ are fermionic fields.

It is worth noting that the explicit verification of the structural identities presented in Ref.~\cite{Brandt:2020vre} remains valid, since the contributions of the additional fields $ \bm{H} $ and $ \bm{\theta} $, $ \bm{\bar{\theta}} $ vanish when dimensional regularization \cite{Leibbrandt:1975dj} is employed. In contrast, Ref.~\cite{Filho:2024kxo} demonstrates that, once these contributions are properly included, the structural identities hold already at the integrand level, independently of the choice of gauge or regularization scheme. This behavior is a direct manifestation of the quantum equivalence between the first- and second-order formulations, rather than a consequence of specific regularization-dependent cancellations.

Moreover, these fields allow computation of the number of degrees of freedom of the first-order EH theory directly from the path integral. Denoting by $N_F$ the number of degrees of freedom associated with a field $F$, the total number of degrees of freedom in the first-order formulation follows straightforwardly:
\begin{equation}\label{eq:dof}
    N^{(1)} = N_{\bm{\phi}^{(1)}} + N_{{H}} - 2 N_{{\theta}}  =N_{\bm{\phi}^{(2)}} + N_{{G}} + N_{{H}} - 2 N_{{\theta}}.
\end{equation}
Note that, $N_{{G} } = N_{H} = N_{\theta} $ (the minus comes from the ghost-like character of the field $ \theta $). Thus, 
\begin{equation}\label{eq:dof1}
    N^{(1)} =  N_{\bm{\phi}^{(2)}} \equiv N^{(2)} =N_{\mathfrak{h}} - 2 N_{d} = 2,
\end{equation}
which correspond to the two physical polarizations of the graviton. This shows that, upon inclusion of the Senjanović determinant, the first- and second-order EH formulations share the same number of physical degrees of freedom.

\subsubsection{Effective action}
Defining the connected generating functional as
$W[j] = -i \ln Z[j]$,
the classical field $\bar{\phi}$ is given by
\begin{equation}\label{eq:defcf}
\bar{\phi}(x) = \frac{\delta W[j]}{\delta j(x)} .
\end{equation}
The effective action $\Gamma[\bar{\phi}]$ is obtained from $W[j]$ by a Legendre transformation,
\begin{equation}\label{eq:defGamma}
    \Gamma[\bar{\phi}] = W[j] - \int \mathop{dx}   j(x)\bar{\phi}(x) .
\end{equation}

In our case, using that $W^{(1)}[ \bm{j} , \bm{J} ] = W^{(2)}[ \bm{j} ; \bm{J} ]$,  we have that 
\begin{equation}\label{eq:defcf12}
    \bar{\mathfrak{h}}^{\mu \nu} 
    = \frac{\delta W^{(1)}[ \bm{j} , \bm{J} ]}{\delta j_{\mu \nu}}
    = \frac{\delta W^{(2)}[ \bm{j} ; \bm{J} ]}{\delta j_{\mu \nu}}
\end{equation}
and 
\begin{equation}\label{eq:defGc}
    \tensor{\bar{G}}{_{\mu \nu}^{\lambda} } 
    =
    \frac{\delta W^{(1)}}{\delta \tensor{J}{^{\mu \nu}_{\lambda}}}
    =
    \frac{\delta W^{(2)}}{\delta \tensor{J}{^{\mu \nu}_{\lambda}}} 
    =
    \tensor{\bar{\mathcal{G}}}{_{\mu \nu}^{\lambda} } [h]
-\kappa^{2} (\bm{M}^{-1} \bm{J}) \tensor{}{_{\mu \nu}^{\lambda} }  
    .
\end{equation}
The effective action of the first-order EH theory is given by 
\begin{equation}\label{eq:defG1}
    \Gamma^{(1)} [ \bar{\mathfrak{h}} , \bar{G} ] = W^{(1)} [\bm{j} , \bm{J} ] - \int \mathop{d x} \left( \bm{j} \cdot \bar{\bm{\mathfrak{h}}} + \bm{J} \cdot \bar{\bm{G}} \right). 
\end{equation}
The effective action corresponding to the generating functional $ Z^{(2)} [ \bm{j} ; \bm{J} ] $ reads 
\begin{equation}\label{eq:defG2}
    \Gamma^{(2)} [ \bar{\mathfrak{h}} ; \bar{G} ] = W^{(2)} [\bm{j} ; \bm{J} ] - \int \mathop{d x} \left( \bm{j} \cdot \bar{\bm{\mathfrak{h}}} + \bm{J} \cdot \bar{\bm{\mathcal{G} }} - \frac{\kappa^{2} }{2} \bm{J} \cdot \bm{M}^{-1} \cdot \bm{J}\right). 
\end{equation}

From Eq.~\eqref{eq:defGc}, we find that 
\begin{equation}\label{eq:relG}
    \Gamma^{(1)} [ \bar{\mathfrak{h}} , \bar{G} ] =
\Gamma^{(2)} [ \bar{\mathfrak{h}} ; \bar{\mathcal{G} } ] + 
\frac{1}{2} 
(\bar{\bm{G}} - \bar{\bm{\mathcal{G} } }  ) 
    \cdot \bm{M} \cdot 
(\bar{\bm{G}} - \bar{\bm{\mathcal{G} } }  ). 
\end{equation}
This relates the one-particle irreducible Green's function of the first- and second-order formulations of the EH theory.
Moreover, when $ \bm{J} = 0$, the effective actions coincide:
\begin{equation}\label{eq:relG2}
\Gamma^{(1)} [ \bar{\mathfrak{h}} , \bar{G} =\bar{\mathcal{G} }( \bar{\mathfrak{h}} ) ] =
    \Gamma^{(2)} [ \bar{\mathfrak{h}} ; \bar{\mathcal{G} } ] \equiv  \Gamma^{(2)} [ \bar{\mathfrak{h}} ],
\end{equation}
where $ \Gamma^{(2)} [ \bar{\mathfrak{h}} ]$ is the effective action of the second-order formulation of the EH theory.
Imposing $ \bm{J} = 0$ in Eq.~\eqref{eq:defGc} implies that the auxiliary field $ \bar{\bm{G}} $ is on-shell. This result was demonstrated for the background effective action in Ref.~\cite{Brandt:2025lkd}.

\subsection{BRST symmetry}
The Faddeev--Popov action is endowed with the well-known BRST symmetry. When we fix a gauge, the initial gauge symmetry is broken. 
However, the gauge symmetry is not entirely lost but survives globally as the BRST symmetry, which reflects the gauge structure of the classical action at the quantum level.

The BRST symmetry of the action \eqref{eq:FP1} reads 
\begin{subequations}\label{eq:BRSTofFOGR}
    \begin{align}
        & \mathsf{s} \mathfrak{h}^{\mu \nu}  = h^{\mu \rho} \partial_{\rho} d^{\nu} -h^{\nu \rho} \partial_{\rho} d^{\mu} - \partial_{\rho} ( h^{\mu \nu} d^{\rho} ) ,\\
        & \mathsf{s} {G}{_{\mu \nu }^{\lambda} }=
    \kappa \bigg[-  \partial_{\mu} \partial_{\nu} d^{\lambda} + \frac{1}{2} ( \delta_{\mu}^{\lambda} \partial_{\nu} + \delta_{\nu}^{\lambda} \partial_{\mu} ) \partial_{\rho} d^{\rho} - d^{\rho} \partial_{\rho} \tensor{G}{_{\mu \nu}^{\lambda}} +   \tensor{G}{_{\mu \nu}^{\rho}} \partial_{\rho} d^{\lambda} - ( \tensor{G}{_{\mu \rho}^{\lambda}} \partial_{\nu} + \tensor{G}{_{\nu \rho}^{\lambda}} \partial_{\mu} )d^{\rho}\bigg],\\
                                                                                                                                                                                                                                                                                                 & \mathsf{s} d^{\mu} = \kappa d^{\rho} \partial_{\rho} d^{\mu}  ,\\
        & \mathsf{s} \bar{d}^{\mu} = B^{\mu} \quad \text{and} \quad \mathsf{s} B^{\mu} =0,
    \end{align}
\end{subequations}
where $ \mathsf{s} $ is the nilpotent BRST operator, $ \mathsf{s}^{2} =0$. 
The invariance of the generating functional \eqref{eq:Z1}, $\mathsf{s} Z^{(1)}=0$, leads to the corresponding Slavnov--Taylor identities. 

Similarly, the same considerations apply to the second-order form, where the BRST symmetry takes the following form \cite{Buchbinder:2021wzv}
\begin{subequations}\label{eq:BRSTofSOGR}
    \begin{align}
        & \mathsf{s} \mathfrak{h}^{\mu \nu}  =  h^{\mu \rho} \partial_{\rho} d^{\nu} -  h^{\nu \rho} \partial_{\rho} d^{\mu}-  \partial_{\rho} (h^{\mu \nu} d^{\rho} ) ,\\
        & \mathsf{s} d^{\mu} = \kappa d^{\rho} \partial_{\rho} d^{\mu}  ,\\
        & \mathsf{s} \bar{d}^{\mu} = B^{\mu} \quad \text{and} \quad  \mathsf{s} B^{\mu} =0.
    \end{align}
\end{subequations}

One can obtain the Slavnov--Taylor identities \cite{Slavnov:1972fg, Taylor:1971ff} using the invariance of the generating functional under the BRST symmetry for both second- and first-order formulations of the EH theory. For completeness, we have included their master equations in the Appendix~B.  

\section{Dyson--Schwinger equation}
We now consider the Dyson--Schwinger equations in the first-order formulation of the EH theory. They lead to interesting identities between the gauge and auxiliary fields. 
Similar identities were derived in Refs.~\cite{McKeon:2020lqp, Brandt:2020vre} using the relation \eqref{eq:qequivgeneral}. 
In contrast, we show that the identities obtained here follow directly from the first-order formulation itself, and independently from the second-order form. This provides a natural interpretation of these structural identities as the quantum realization of the classical equations of motion.

The Dyson--Schwinger equations \cite{Dyson:1949ha, Schwinger:1951ex} follow from the invariance of the generating functional under infinitesimal changes of the integration variables, corresponding to functional integration by parts \cite{itzyksonQuantumFieldTheory1980}. 
We now implement this ``functional integration by parts'' in the expectation value\footnote{Since the Faddeev--Popov (FP) sector of the quantum action is independent of the auxiliary field, it does not contribute to the Dyson--Schwinger equation. Therefore, only the classical action $S^{(1)}$ enters.}
\begin{equation}\label{eq:derivation1}
    \left\langle 0\left|T \frac{\delta \mathcal{O}_{i}( x_{i} )}{\delta \bm{G}(x)}  \right| 0 \right\rangle^{(1)}
    = -i \left\langle 0\left|T \mathcal{O}_{i} ( x_{i} ) \frac{\delta S^{(1)} }{\delta \bm{G} (x)}\right| 0 \right\rangle^{(1)},
\end{equation}
where we use the notation
\begin{equation}\label{eq:defexpv}
    \langle 0|T \mathcal{O}( \ldots)  | 0 \rangle^{(1)}_{\bm{j} , \bm{J}} \equiv 
\int \mathcal{D} {\bm{\phi}}^{(1)}
    |\det \bm{M}(h) |^{1/2}
    \mathcal{O} (\ldots) 
    \exp\left[i\left(S_{\text{FP}}^{(1)}[{\bm{\phi}}^{(1)}] + \int \mathop{d x} \left(\bm{j}  \cdot \bm{\mathfrak{h} }  + \bm{J}  \cdot \bm{G} \right)\right)\right].
\end{equation}

Using $ \mathcal{O}_{i} (x_{i}) \to  \kappa^{2} \mathcal{O}_{i} (x_{i} ) \cdot \bm{M}^{-1}  (h)$, Eq.~\eqref{eq:derivation1} becomes
\begin{equation}\label{eq:derivation2}
\kappa^{2} \left\langle 0\left|T  \frac{\delta \mathcal{O}_{i}( x_{i} )}{\delta \bm{G}(x)}  \cdot \bm{M}^{-1} \right| 0 \right\rangle^{(1)}
= -i \kappa^{2} \left\langle 0\left|T \mathcal{O}_{i} ( x_{i} ) \bm{M}^{-1} \cdot \frac{\delta S^{(1)} }{\delta \bm{G} (x)}\right| 0 \right\rangle^{(1)}.
\end{equation}

Substituting the explicit form of $ S^{(1)} $, we obtain 
\begin{equation}\label{eq:derivation3}
\left\langle 0\left|T \mathcal{O}_{i} ( x_{i} ) \left[G - \mathcal{G} \right] \tensor{}{_{\mu \nu}^{\lambda}}(x)  \right| 0 \right\rangle^{(1)}
    = i \kappa^{2} 
    \left\langle 0\left|T  \frac{\delta \mathcal{O}_{i}( x_{i} )}{\delta \tensor{G}{_{\mu' \nu'}^{\lambda'}} (x)}   \tensor{({M}^{-1})}{_{\mu ' \nu '}^{\lambda '}_{\mu \nu}^{\lambda} } \right| 0 \right\rangle^{(1)}.
\end{equation}

It is thus clear that the Dyson--Schwinger equation~\eqref{eq:derivation3} corresponds to the quantum counterpart of the classical equation of motion
\begin{equation}\label{eq:eomclassical}
    \tensor{G}{_{\mu \nu}^{\lambda}} = \tensor{\mathcal{G} }{_{\mu \nu}^{\lambda} } [h ]. 
\end{equation}
In particular, for $ \mathcal{O}_{1} = 1$, we obtain
\begin{equation}\label{eq:sid00}
    \langle 0|T \tensor{G}{_{\mu \nu}^{\lambda}} (x)| 0 \rangle^{(1)} 
    = \langle 0|T \tensor{\mathcal{G}}{_{\mu \nu}^{\lambda}} (x)| 0 \rangle^{(1)} .
\end{equation}

We can use the Dyson--Schwinger equation~\eqref{eq:derivation3} to derive structural identities, such as Eq.~\eqref{eq:SID}, which we refer to as \emph{intrinsic structural identities}. 
Additional identities can be obtained by considering different choices of operators $\mathcal{O}_i$ in Eq.~\eqref{eq:derivation3}, thereby generating a hierarchy of interdependent constraints. 
In contrast to the structural identities obtained in Ref.~\cite{Brandt:2020vre}, these identities arise directly from the first-order formulation of the EH theory.

As an example, we derive the intrinsic structural identity shown in Eq.~\eqref{eq:SID}. 
For $ \mathcal{O}_{1} (y) = \bm{G} (y)$, Eq.~\eqref{eq:derivation3} gives
\begin{equation}\label{eq:d3intosid2}
    \langle 0|T \tensor{(G - \mathcal{G} )}{_{\mu \nu}^{\lambda} }(x) \tensor{G}{_{\pi \tau}^{\gamma}} (y) | 0 \rangle 
    = i \kappa^{2} \delta (x-y) \langle 0|T \tensor{(M^{-1})}{_{\mu \nu}^{\lambda}_{\pi \tau}^{\gamma}}| 0 \rangle^{(1)} . 
\end{equation}

This implies
\begin{equation}\label{eq:sid1part1}
\langle 0|T \tensor{G}{_{\mu \nu}^{\lambda}} (x) \tensor{G}{_{\pi \tau}^{\gamma} } (y)| 0 \rangle^{(1)} 
= i \kappa^{2} \delta (x-y)\langle 0|T \tensor{(M^{-1})}{_{\mu \nu}^{\lambda}_{\pi \tau}^{\gamma}}(x)| 0 \rangle^{(1)} 
+  \langle 0|T \tensor{\mathcal{G}}{_{\mu \nu}^{\lambda}} (x) \tensor{{G}}{_{\pi \tau}^{\gamma} } (y)| 0 \rangle^{(1)}.
\end{equation}

Next, taking $ \mathcal{O}_{1} (x_{1} ) = \bm{\mathcal{G} } (y)$ in Eq.~\eqref{eq:derivation3}, we obtain
\begin{equation}\label{eq:sid1part2}
    \langle 0|T \tensor{G}{_{\mu \nu}^{\lambda}} (x) \tensor{\mathcal{G}}{_{\pi \tau}^{\gamma} } (y)| 0 \rangle^{(1)} 
    = 
    \langle 0|T \tensor{\mathcal{G}}{_{\mu \nu}^{\lambda}} (x) \tensor{{\mathcal{G} }}{_{\pi \tau}^{\gamma} } (y)| 0 \rangle^{(1)}.
\end{equation}

Substituting Eq.~\eqref{eq:sid1part2} into Eq.~\eqref{eq:sid1part1} leads to the intrinsic structural identity \eqref{eq:SID1}. 
The identity~\eqref{eq:SID2} is obtained analogously by taking $ \mathcal{O}_{1} = \bm{\mathfrak{h}} (y)$ in Eq.~\eqref{eq:derivation3}.

Let us establish a connection between the intrinsic structural identities obtained here and those derived in Ref.~\cite{Brandt:2020vre}. Using the quantum equivalence relation \eqref{eq:qequivgeneral}, we obtain
\begin{equation}\label{eq:derivation5a}
    \frac{\delta}{\delta \tensor{J}{^{\mu \nu}_{\lambda}} (x) } \left[\left\langle 0 \left|T  \right| 0 \right\rangle_{\bm{j} , \bm{J}}^{(1)} 
    -
\left\langle 0 \left|T  \right| 0 \right\rangle_{\bm{j} ;\bm{J}}^{(2)}\right] = 0. 
\end{equation}
Since Green's functions of the auxiliary field $ \bm{\mathcal{G}} $ are functions of the metric $ \bm{\mathfrak{h}} $, we can use the relation \eqref{eq:equivh}, which implies that 
\begin{equation}\label{eq:derivation5b}
    \frac{\delta}{\delta \tensor{J}{^{\mu \nu}_{\lambda}} (x) }\left\langle 0 \left|T  \right| 0 \right\rangle_{\bm{j} ;\bm{J}}^{(2)} =     
    \frac{\delta}{\delta \tensor{J}{^{\mu \nu}_{\lambda}} (x) }\left\langle 0 \left|T  \right| 0 \right\rangle_{\bm{j} ;\bm{J}}^{(1)}.  
\end{equation}
Combining these results, we obtain
\begin{equation}\label{eq:derivation5c}
    \frac{\delta}{\delta \tensor{J}{^{\mu \nu}_{\lambda}} (x) } \left[\left\langle 0 \left|T  \right| 0 \right\rangle_{\bm{j} , \bm{J}}^{(1)} 
    -
\left\langle 0 \left|T  \right| 0 \right\rangle_{\bm{j} ;\bm{J}}^{(1)}\right] = 0,
\end{equation}
which leads to the structural identities \eqref{eq:SID10} and \eqref{eq:SID20}, as well as their higher-order generalizations.

It is important to emphasize that Eq.~\eqref{eq:derivation5c} relates functional derivatives of two distinct generating functionals, corresponding to different realizations of the auxiliary sector. As a consequence, the resulting identities involve both composite operators and explicit source-dependent contributions.
In Eq.~\eqref{eq:derivation5c}, one encounters two different functional derivatives:
\begin{equation}\label{eq:realtios}
    \frac{\delta \langle 0|T| 0 \rangle^{(1)}_{\bm{j} , \bm{J}}}{\delta J}
    =
    i \langle 0|T \bm{G}| 0 \rangle^{(1)}_{\bm{j} , \bm{J}}
    \quad \text{and} \quad  
    \frac{\delta \langle 0|T| 0 \rangle^{(1)}_{\bm{j} ; \bm{J}}}{\delta J}
    =
    i \left\langle 0\left|T \bm{\mathcal{G}} (h)- \bm{M}^{-1} \cdot \bm{J}\right| 0 \right\rangle_{\bm{j} ; \bm{J}}^{(1)},
\end{equation}
which in both the Green's function in the first-order formulation of the EH theory.

\section{Conclusions}

In this paper, we have studied the covariant quantization of the first-order formulation of EH gravity. We analyzed the gauge algebra of both first- and second-order formulations, showing that they are closed and irreducible. 
We showed that a novel local symmetry appears in the first-order formulation of the EH theory. However, similarly to the Yang--Mills symmetry found in Ref.~\cite{Lavrov:2021pqh}, this symmetry is trivial and therefore does not affect its gauge algebra. 

Using the BV formalism, we constructed the quantum action and derived a manifestly covariant expression for the Senjanović determinant, thereby avoiding the non-covariant measures of earlier approaches. 
This, in turn, allows us to establish the exact quantum equivalence between the first- and second-order formulations at the level of the generating functional $Z$ and the effective action $\Gamma$.

We then examined the Dyson--Schwinger equations of the EH theory in first-order form and showed that they generate a family of structural identities [see, for example, Eqs.~\eqref{eq:SID}] that constrain the Green's functions of the auxiliary field. 
Although they can be related to previously derived identities through~\eqref{eq:qequivgeneral}, they emerge intrinsically within the first-order formulation of the EH theory.
Moreover, our results clarify the interpretation of such identities at the quantum level. 
In the present framework, the intrinsic structural identities follow directly from the Dyson--Schwinger equation \eqref{eq:derivation3} and can thus be identified with the quantum realization of the classical equations of motion~\eqref{eq:eomclassical}.

Looking ahead, the ideas developed here may be extended to supergravity, where first-order formulations arise naturally \cite{VanNieuwenhuizen:1981ae}, and to theories with more complicated auxiliary field structures \cite{Ferko:2024ali, Fukushima:2024nxm}. The structural identities may provide insight into nonperturbative aspects of quantum gravity, since they are exact relations between field correlations. Moreover, the role of the Senjanović determinant in curved spacetime and at finite temperature also deserves further investigation.

\begin{acknowledgments}
    We thank J.\ Frenkel, F.\ T.\ Brandt, D.\ G.\ C.\ McKeon for enlightening conversations.
S.\ M.-F. thanks FAPESP for partial financial support.
This study was financed, in part, by the São Paulo Research Foundation (FAPESP), Brasil. Process Number \#2025/16156-7. 
\end{acknowledgments}

\appendix
\section{Senjanović determinant in finite temperature} \label{sec:finite}

\begin{figure}[ht]
    \includegraphics[scale=0.8]{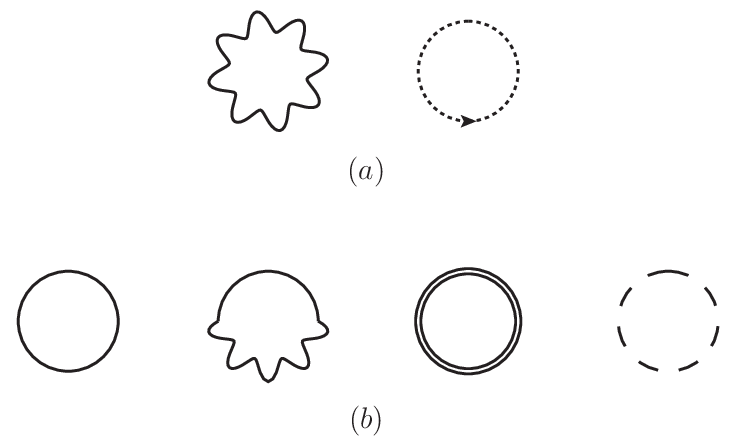}
    \caption{Diagrams (a) are the one-loop contribution to the free energy in the second-order formulation of the EH theory. Diagrams (b) are the contributions that arise in the first-order EH theory. Wavy lines represent the graviton field $ \mathbf{\mathfrak{h}} $, dotted lines are the Faddeev--Popov ghost fields $ \bm{\bar{d}} $, $ \bm{d} $ and solid lines denote the auxiliary field $ \mathbf{G} $. The doubled solid, dashed lines denote respectively the fields $ \mathbf{H} $, $ \bar{\bm{\theta}} $ and $ \bm{\theta} $, which arise from the Senjanović determinant $ \det^{1/2} \bm{M} (h)$.}\label{fig:temp}
\end{figure}

Consider the one-loop contributions to the free energy at finite temperature shown in Fig.~\ref{fig:temp}. 
In the second-order formulation of the EH theory, we have:
\begin{equation}\label{eq:contrT}
    F_{\text{one-loop}}^{(2)} 
\equiv - \ln Z^{(2)} \big|_{\text{one-loop}}  = 
\frac{1}{2}     
\mathop{\rm Tr} \ln \frac{\delta^{2} S^{(2)}_{\text{FP}}}{\delta \bm{\mathfrak{h}} \delta \bm{\mathfrak{h}}} \bigg|_{ \phi^{(2)} =0} 
    -
    \mathop{\rm Tr} \ln 
\frac{\delta^{2} S^{(2)}_{\text{FP}}}{\delta \bm{\bar{d}} \delta \bm{d}} 
\bigg|_{ \phi^{(2)} =0}.
\end{equation}

In the first-order formulation, it reads:
\begin{equation}\label{eq:contrT2}
    F_{\text{one-loop}}^{(1)} 
=
\frac{1}{2} \mathop{\rm Tr} \ln \frac{\delta^{2} S^{(1)}_{\text{FP}}}{\delta \bm{\hat{F}} \delta \bm{\hat{F}}} \bigg|_{ \phi^{(1)} =0} 
    -
    \mathop{\rm Tr} \ln 
\frac{\delta^{2} S^{(1)}_{\text{FP}}}{\delta \bm{\bar{d}} \delta \bm{d}} 
\bigg|_{ \phi^{(1)} =0}
- \frac{1}{2} \mathop{\rm Tr} \ln \bm{M} (h) ,
\end{equation}
where the last term arises from the Senjanović determinant and we have defined 
$\bm{\hat{F}} = \begin{pmatrix} \bm{\mathfrak{h}} & \bm{G} \end{pmatrix}$.

Diagonalizing the first term on the right-hand side of Eq.~\eqref{eq:contrT2} gives
\begin{equation}\label{eq:new}
\mathop{\rm Tr} \ln \frac{\delta^{2} S^{(1)}_{\text{FP}}}{\delta \bm{\hat{F}} \delta \bm{\hat{F}}} \bigg|_{ \phi^{(1)} =0} =
\mathop{\rm Tr} \ln \frac{\delta^{2} S^{(2)}_{\text{FP}}}{\delta \bm{\mathfrak{h}} \delta \bm{\mathfrak{h}}} \bigg|_{ \phi^{(2)} =0} 
+ 
\mathop{\rm Tr} \ln \frac{\delta^{2} S^{(1)}_{\text{FP}}}{\delta \bm{G} \delta \bm{G}} \bigg|_{ \phi^{(1)} =0}.
\end{equation}

Using this in Eq.~\eqref{eq:contrT2}, we then obtain that
\begin{equation}\label{eq:equivF}
    F^{(2)}_{\text{one-loop}} = F^{(1)}_{\text{one-loop}},
\end{equation}
since the ghost sector is the same in both formulations. We see that the Senjanović determinant cancels the auxiliary field $ \bm{G} $ contributions to the free energy at finite temperature. This illustrates the essential role of the Senjanović determinant in ensuring the quantum equivalence between the two formulations.

\section{Master equations for the BRST symmetry}\label{sec:B}
In this Appendix, we show the master equation for the BRST symmetries of the second- \eqref{eq:BRSTofSOGR} and first-order Eq.~\eqref{eq:BRSTofFOGR} formulations of the EH theory. 

\subsection{Second-order formulation}
 
We obtain the Slavnov--Taylor identities by extending the source term. In this formulation, the extended source term is given by
\begin{equation}\label{eq:sourceof2GRextended}
    j_{\mu \nu} \mathfrak{h}^{\mu \nu} + i( \bar{\eta}_{\mu} d^{\mu} - \bar{d}^{\mu} \eta_{\mu} ) 
    + j_{\mu} B^{\mu}
    +  K_{\mu \nu}\left[ 
 h^{\mu \rho} \partial_{\rho} d^{\nu} +  h^{\nu \rho} \partial_{\rho} d^{\mu}-  \partial_{\rho} (h^{\mu \nu} d^{\rho} ) 
\right]
+  \kappa K_{\mu} d^{\rho} \partial_{\rho} d^{\mu}.
\end{equation}
Note that, we have added sources to the non-linear BRST variations, these sources in the BV formalism are the antifields.

The corresponding Zinn-Justin master equation is given by \cite{Filho:2024kxo}
\begin{equation}\label{eq:MSofSOGR}
    {\left ( \Gamma^{(2)} , \Gamma^{(2)}\right )} = 
    \int \mathop{d x} \left (
        \Gamma^{(2)}\frac{\overset{\leftarrow}{\delta}  }{\delta \mathfrak{h}^{\mu \nu} } 
        \frac{\overset{\rightarrow}{\delta}  \Gamma^{(2)}}{\delta K_{\mu \nu}} + 
    \Gamma^{(2)}\frac{\overset{\leftarrow}{\delta}  }{\delta d^{\mu }} \frac{\overset{\rightarrow}{\delta} \Gamma^{(2)} }{\delta K_{\mu }}  + \Gamma^{(2)}\frac{\overset{\leftarrow}{\delta}  }{\delta \bar{d}^{\mu }} \frac{\overset{\rightarrow}{\delta} \Gamma^{(2)} }{\delta j_{\mu }}\right) =0.
\end{equation}
The Slavnov--Taylor identities of the second-order formulation of gravity are derived by taking functional derivatives of the master equation~\eqref{eq:MSofSOGR}.

\subsection{First-order formulation}

In the first-order form, the extended source term reads 
\begin{equation}\label{eq:sourceof1GRextended}
    \begin{split}
    & j^{\mu \nu} \mathfrak{h}_{\mu \nu} + {J}{_{\lambda}^{\mu \nu}} \tensor{G}{_{\mu \nu}^{\lambda}} + i( \bar{\eta}_{\mu} d^{\mu} - \bar{d}^{\mu} \eta_{\mu} ) 
+ j_{\mu} B^{\mu}
 \\ & 
+K_{\mu \nu } \left[
h^{\mu \rho} \partial_{\rho} d^{\nu} -h^{\nu \rho} \partial_{\rho} d^{\mu} - \partial_{\rho} ( h^{\mu \nu} d^{\rho} ) 
\right]
+ \kappa  K_{\mu} d^{\rho} \partial_{\rho } d^{\mu} 
\\ & 
+ \kappa \tensor{K}{^{\mu \nu}_{\lambda}} 
        \bigg[- \partial_{\mu} \partial_{\nu} d^{\lambda} + \frac{1}{2} ( \delta_{\mu}^{\lambda} \partial_{\nu} + \delta_{\nu}^{\lambda} \partial_{\mu} ) \partial_{\rho} d^{\rho} - d^{\rho} \tensor{G}{_{\mu \nu}^{\lambda}} 
+ \tensor{G}{_{\mu \nu}^{\rho}} \partial_{\rho} d^{\lambda} - ( \tensor{G}{_{\mu \rho}^{\lambda}} \partial_{\nu} + \tensor{G}{_{\nu \rho}^{\lambda}} \partial_{\mu} )d^{\rho} \bigg].
    \end{split}
\end{equation}
The corresponding Zinn-Justin master equation is given by
\begin{equation}\label{eq:MSofFOGR}
    {\left ( \Gamma^{(1)} , \Gamma^{(1)}\right )} = 
    \int \mathop{d x} \left (
        \Gamma^{(1)}\frac{\overset{\leftarrow}{\delta}  }{\delta \mathfrak{h}^{\mu \nu} } \frac{\overset{\rightarrow}{\delta} \Gamma^{(1)} }{\delta K_{\mu \nu}} + 
        \Gamma^{(1)}\frac{\overset{\leftarrow}{\delta}  }{\delta \tensor{G}{_{\mu \nu}^{\lambda}} } \frac{\overset{\rightarrow}{\delta} \Gamma^{(1)} }{\delta \tensor{K}{^{\mu \nu}_{\lambda }}} + 
    \Gamma^{(1)}\frac{\overset{\leftarrow}{\delta}  }{\delta d^{\mu }} \frac{\overset{\rightarrow}{\delta} \Gamma^{(1)} }{\delta K_{\mu }}  + \Gamma^{(1)}\frac{\overset{\leftarrow}{\delta}  }{\delta \bar{d}^{\mu }} \frac{\overset{\rightarrow}{\delta} \Gamma^{(1)} }{\delta j_{\mu }}\right) =0.
\end{equation}
The Slavnov--Taylor identities of the first-order formulation of gravity are derived by taking functional derivatives of the master equation~\eqref{eq:MSofFOGR}.

\bibliography{covquantization.bib}

@article{Nakanishi:1966zz,
  title = {Covariant {{Quantization}} of the {{Electromagnetic Field}} in the {{Landau Gauge}}},
  author = {Nakanishi, Noboru},
  year = 1966,
  journal = {Progress of Theoretical Physics},
  volume = {35},
  number = {RIMS-13},
  pages = {1111--1116},
  doi = {10.1143/PTP.35.1111}
}

@article{Faddeev:1967fc,
  title = {Feynman {{Diagrams}} for the {{Yang-Mills Field}}},
  author = {Faddeev, L.D. and Popov, V.N.},
  year = 1967,
  journal = {Physics Letters B},
  volume = {25},
  number = {1},
  pages = {29--30},
  issn = {03702693},
  doi = {10.1016/0370-2693(67)90067-6},
  urldate = {2022-06-20},
  langid = {english}
}

@article{lautrup:1967,
  title = {Canonical Quantum Electrodynamics in Covariant Gauges},
  author = {Lautrup, B.},
  year = 1967,
  journal = {Kong. Dan. Vid. Sel. Mat. Fys. Med.},
  volume = {35},
  number = {11}
}

@article{Tyutin:1975qk,
  title = {Gauge {{Invariance}} in {{Field Theory}} and {{Statistical Physics}} in {{Operator Formalism}}},
  author = {Tyutin, I. V.},
  year = 1975,
  journal = {P. N. Lebedev Physical Institute},
  volume = {39},
  number = {LEBEDEV-75-39},
  doi = {10.48550/arXiv.0812.0580}
}

@article{buchbinder:1983a,
  title = {One {{Loop Calculation}} of the {{Polarization Operator}} of {{Gravitons}} in the {{First Order Formalism}}.},
  author = {{l. Buchbinder}, I. and {l. Shapiro}, I.},
  year = 1983,
  journal = {Soviet Journal of Nuclear Physics},
  volume = {37},
  pages = {248--252}
}

@article{buchbinder:1985,
  title = {One {{Loop Counterterms Derivation}} in the {{First Order Quantum Gravity Formalism}}},
  author = {Buchbinder, I. L. and Shapiro, I. L.},
  year = 1985,
  journal = {Acta Physica Polonica B},
  volume = {16},
  pages = {103--107},
  doi = {https://www.actaphys.uj.edu.pl/R/16/2/103/pdf}
}

@article{McKeon:1994ds,
  title = {The {{First}} Order Formalism for {{Yang-Mills}} Theory},
  author = {McKeon, D. G. C.},
  year = 1994,
  journal = {Canadian Journal of Physics},
  volume = {72},
  pages = {601--607},
  doi = {10.1139/p94-077}
}

@article{McKeon:2010nf,
  title = {The {{Canonical Structure}} of the {{First Order Einstein-Hilbert Action}}},
  author = {McKeon, D. G. C.},
  year = 2010,
  journal = {International Journal of Modern Physics A},
  volume = {25},
  number = {17},
  eprint = {1005.3001},
  primaryclass = {gr-qc},
  pages = {3453--3480},
  issn = {0217-751X, 1793-656X},
  doi = {10.1142/S0217751X10050093},
  urldate = {2021-10-10},
  archiveprefix = {arXiv},
  langid = {english}
}

@article{Kiriushcheva:2006gp,
  title = {A {{Canonical Analysis}} of the {{Einstein-Hilbert Action}} in {{First Order Form}}},
  author = {Kiriushcheva, N. and Kuzmin, S.V. and McKeon, D. G. C.},
  year = 2006,
  journal = {International Journal of Modern Physics A},
  volume = {21},
  eprint = {hep-th/0609219},
  pages = {3401--3420},
  doi = {10.1142/S0217751X06029545},
  archiveprefix = {arXiv}
}

@article{Leibbrandt:1975dj,
  title = {Introduction to the {{Technique}} of {{Dimensional Regularization}}},
  author = {Leibbrandt, George},
  year = 1975,
  journal = {Reviews of Modern Physics},
  volume = {47},
  number = {4},
  pages = {849},
  issn = {0034-6861},
  doi = {10.1103/RevModPhys.47.849},
  urldate = {2020-07-14},
  langid = {english}
}

@article{Okubo:1979gt,
  title = {Duffin-Kemmer {{Formulation}} of {{Gauge Theories}}},
  author = {Okubo, Susumu and Tosa, Yasunari},
  year = 1979,
  journal = {Physical Review D},
  volume = {20},
  number = {2},
  pages = {462},
  issn = {0556-2821},
  doi = {10.1103/PhysRevD.20.462},
  urldate = {2020-08-28},
  langid = {english}
}

@article{Brandt:2020vre,
  title = {Structural Identities in the First-Order Formulation of Quantum Gravity},
  author = {Brandt, F. T. and Frenkel, J. and {Martins-Filho}, S. and McKeon, D. G. C.},
  year = 2020,
  month = aug,
  journal = {Physical Review D},
  volume = {102},
  number = {4},
  eprint = {2007.04841},
  primaryclass = {hep-th},
  pages = {045013},
  publisher = {American Physical Society},
  doi = {10.1103/PhysRevD.102.045013},
  archiveprefix = {arXiv},
  copyright = {All rights reserved}
}

@article{Brandt:2016eaj,
  title = {Radiative {{Corrections}} and the {{Palatini Action}}},
  author = {Brandt, F.T. and McKeon, D.G.C.},
  year = 2016,
  month = may,
  journal = {Physical Review D},
  volume = {93},
  number = {10},
  eprint = {1601.04944},
  primaryclass = {hep-th},
  pages = {105037},
  doi = {10.1103/PhysRevD.93.105037},
  archiveprefix = {arXiv}
}

@article{Lavrov:2021pqh,
  title = {Covariant Quantization of {{Yang-Mills}} Theory in the First Order Formalism},
  author = {Lavrov, P.M.},
  year = 2021,
  month = may,
  journal = {Physics Letters B},
  volume = {816},
  eprint = {2101.06868},
  primaryclass = {hep-th},
  pages = {136182},
  issn = {03702693},
  doi = {10.1016/j.physletb.2021.136182},
  urldate = {2021-10-27},
  archiveprefix = {arXiv},
  langid = {english}
}

@article{McKeon:2020lqp,
  title = {Consistency {{Conditions}} for the {{First-Order Formulation}} of {{Yang-Mills Theory}}},
  author = {Brandt, F. T. and Frenkel, J. and {Martins-Filho}, S. and McKeon, D. G. C.},
  year = 2020,
  month = apr,
  journal = {Physical Review D},
  volume = {101},
  number = {8},
  eprint = {2003.06819},
  primaryclass = {hep-th},
  pages = {085013},
  doi = {10.1103/PhysRevD.101.085013},
  archiveprefix = {arXiv}
}

@article{Brandt:2015nxa,
  title = {Perturbative {{Calculations}} with the {{First Order Form}} of {{Gauge Theories}}},
  author = {Brandt, F.T. and McKeon, D.G.C.},
  year = 2015,
  month = may,
  journal = {Physical Review D},
  volume = {91},
  number = {10},
  eprint = {1503.02598},
  primaryclass = {hep-th},
  pages = {105006},
  issn = {1550-7998, 1550-2368},
  doi = {10.1103/PhysRevD.91.105006},
  urldate = {2023-02-02},
  archiveprefix = {arXiv},
  langid = {english}
}

@article{Ferraris:1982wci,
  title = {Variational Formulation of General Relativity from 1915 to 1925 ``{{Palatini}}'s Method'' Discovered by {{Einstein}} in 1925},
  author = {Ferraris, M. and Francaviglia, M. and Reina, C.},
  year = 1982,
  month = mar,
  journal = {Gen. Rel. Grav.},
  volume = {14},
  number = {3},
  pages = {243--254},
  doi = {10.1007/BF00756060}
}

@article{Goldberg:1958zz,
  title = {Conservation {{Laws}} in {{General Relativity}}},
  author = {Goldberg, J.N.},
  year = 1958,
  journal = {Physical Review},
  volume = {111},
  pages = {315--320},
  doi = {10.1103/PhysRev.111.315}
}

@article{VanNieuwenhuizen:1981ae,
  title = {Supergravity},
  author = {Van Nieuwenhuizen, P.},
  year = 1981,
  journal = {Physics Reports},
  volume = {68},
  pages = {189--398},
  doi = {10.1016/0370-1573(81)90157-5}
}

@article{Brandt:2024rsy,
  title = {Quantization of {{Einstein-Cartan}} Theory in the First Order Form},
  author = {Brandt, F. T. and Frenkel, J. and {Martins-Filho}, S. and McKeon, D. G. C.},
  year = 2024,
  month = jan,
  journal = {Annals of Physics},
  volume = {462},
  eprint = {2401.16343},
  primaryclass = {hep-th},
  pages = {169607},
  issn = {00034916},
  doi = {10.1016/j.aop.2024.169607},
  urldate = {2024-01-29},
  archiveprefix = {arXiv},
  langid = {english}
}

@phdthesis{Filho:2024kxo,
  title = {Covariant Quantization of Gauge Theories with {{Lagrange}} Multipliers},
  author = {{Martins-Filho}, S.},
  year = 2024,
  month = dec,
  eprint = {2504.04666},
  primaryclass = {hep-th},
  address = {S\~ao Paulo},
  doi = {10.48550/arXiv.2504.04666},
  urldate = {2023-01-23},
  archiveprefix = {arXiv},
  langid = {english},
  school = {Universidade de S\~ao Paulo}
}

@article{Brandt:2024kvs,
  title = {Renormalization of the {{Einstein-Cartan Theory}} in {{First-Order Form}}},
  author = {Brandt, F. T. and Frenkel, J. and {Martins-Filho}, S. and McKeon, D. G. C.},
  year = 2024,
  month = sep,
  journal = {Annals of Physics},
  volume = {470},
  eprint = {2409.10493},
  primaryclass = {hep-th},
  pages = {169801},
  issn = {00034916},
  doi = {10.1016/j.aop.2024.169801},
  urldate = {2024-01-29},
  archiveprefix = {arXiv},
  langid = {english}
}

@article{Aros:2003bi,
  title = {Path Integral Measure for First Order and Metric Gravities},
  author = {Aros, Rodrigo and Contreras, Mauricio and Zanelli, Jorge},
  year = 2003,
  journal = {Classical and Quantum Gravity},
  volume = {20},
  number = {13},
  eprint = {gr-qc/0303113},
  pages = {2937--2944},
  issn = {0264-9381, 1361-6382},
  doi = {10.1088/0264-9381/20/13/336},
  urldate = {2025-11-27},
  archiveprefix = {arXiv},
  langid = {english}
}

@article{Batalin:1981jr,
  title = {Gauge {{Algebra}} and {{Quantization}}},
  author = {Batalin, I.A. and Vilkovisky, G.A.},
  year = 1981,
  journal = {Second Seminar on Quantum Gravity Moscow, USSR, October 13-15, 1981},
  volume = {102},
  pages = {27--31},
  doi = {10.1016/0370-2693(81)90205-7}
}

@article{Brandt:2020gms,
  title = {On Restricting First Order Form of Gauge Theories to One-Loop Order},
  author = {Brandt, F. T. and Frenkel, J. and {Martins-Filho}, S. and McKeon, D. G. C.},
  year = 2021,
  month = apr,
  journal = {Annals of Physics},
  volume = {427},
  eprint = {2009.09553},
  primaryclass = {hep-th},
  pages = {168426},
  issn = {00034916},
  doi = {10.1016/j.aop.2021.168426},
  urldate = {2022-07-08},
  archiveprefix = {arXiv},
  copyright = {All rights reserved},
  langid = {english}
}

@article{Batalin:1983ggl,
  title = {Quantization of {{Gauge Theories}} with {{Linearly Dependent Generators}}},
  author = {Batalin, I.A. and Vilkovisky, G.A.},
  year = 1983,
  journal = {Physical Review D},
  volume = {28},
  number = {10},
  pages = {2567--2582},
  issn = {0556-2821},
  doi = {10.1103/PhysRevD.28.2567},
  urldate = {2023-12-10},
  langid = {english}
}

@book{Buchbinder:2021wzv,
  title = {Introduction to Quantum Field Theory with Applications to Quantum Gravity},
  author = {Buchbinder, Iosif L. and Shapiro, Ilya L.},
  year = 2021,
  series = {Oxford Graduate Texts},
  edition = {First edition},
  publisher = {Oxford University Press},
  address = {Oxford},
  doi = {10.1093/oso/9780198838319.001.0001},
  urldate = {2026-03-04},
  isbn = {978-0-19-883831-9},
  langid = {english}
}

@article{Lavrov:1997xk,
  title = {Quantization of Two-Dimensional Gravity with Dynamical Torsion},
  author = {Lavrov, P. M. and Moshin, P. {\relax Yu}.},
  year = 1999,
  journal = {Classical and Quantum Gravity},
  volume = {16},
  number = {7},
  eprint = {hep-th/9710100},
  pages = {2247--2258},
  issn = {0264-9381, 1361-6382},
  doi = {10.1088/0264-9381/16/7/307},
  urldate = {2025-03-27},
  archiveprefix = {arXiv},
  langid = {english}
}

@article{Kallosh:1978de,
  title = {Modified {{Feynman Rules}} in {{Supergravity}}},
  author = {Kallosh, R.E.},
  year = 1978,
  journal = {Nuclear Physics B},
  volume = {141},
  number = {1-2},
  pages = {141--152},
  issn = {05503213},
  doi = {10.1016/0550-3213(78)90340-1},
  urldate = {2022-06-20},
  langid = {english}
}

@article{Nielsen:1978mp,
  title = {Ghost {{Counting}} in {{Supergravity}}},
  author = {Nielsen, N.K.},
  year = 1978,
  journal = {Nuclear Physics B},
  volume = {140},
  number = {3},
  pages = {499--509},
  issn = {05503213},
  doi = {10.1016/0550-3213(78)90009-3},
  urldate = {2022-06-20},
  langid = {english}
}

@article{Senjanovic:1976br,
  title = {Path {{Integral Quantization}} of {{Field Theories}} with {{Second Class Constraints}}},
  author = {Senjanovic, P.},
  year = 1976,
  journal = {Annals of Physics},
  volume = {100},
  number = {1-2},
  pages = {227--261},
  issn = {00034916},
  doi = {10.1016/0003-4916(76)90062-2},
  urldate = {2021-10-09},
  langid = {english}
}

@article{Stelle:1976gc,
  title = {Renormalization of {{Higher Derivative Quantum Gravity}}},
  author = {Stelle, K.S.},
  year = 1977,
  journal = {Physical Review},
  volume = {16},
  pages = {953--969},
  doi = {10.1103/PhysRevD.16.953}
}

@article{Schwinger:1951ex,
  title = {On the {{Green}}'s Functions of Quantized Fields. 1.},
  author = {Schwinger, Julian S.},
  year = 1951,
  journal = {Proc. Nat. Acad. Sci.},
  volume = {37},
  number = {7},
  pages = {452--455},
  issn = {0027-8424, 1091-6490},
  doi = {10.1073/pnas.37.7.452},
  urldate = {2026-04-27},
  langid = {english}
}

@article{Becchi:1974xu,
  title = {The {{Abelian Higgs-Kibble Model}}. {{Unitarity}} of the {{S Operator}}},
  author = {Becchi, C. and Rouet, A. and Stora, R.},
  year = 1974,
  journal = {Physics Letters B},
  volume = {52},
  number = {3},
  pages = {344--346},
  doi = {10.1016/0370-2693(74)90058-6}
}

@article{Dyson:1949ha,
  title = {The {{S}} Matrix in Quantum Electrodynamics},
  author = {Dyson, F.J.},
  year = 1949,
  journal = {Physical Review},
  volume = {75},
  number = {11},
  pages = {1736--1755},
  issn = {0031-899X},
  doi = {10.1103/PhysRev.75.1736},
  urldate = {2026-04-27},
  copyright = {http://link.aps.org/licenses/aps-default-license},
  langid = {english}
}

@article{Chishtie:2011wd,
  title = {Non-{{Trivial Ghosts}} and {{Second Class Constraints}}},
  author = {Chishtie, Farrukh and McKeon, D.G.C.},
  year = 2012,
  journal = {International Journal of Modern Physics A},
  volume = {27},
  number = {14},
  eprint = {1110.1425},
  primaryclass = {hep-th},
  pages = {1250077},
  issn = {0217-751X, 1793-656X},
  doi = {10.1142/S0217751X12500777},
  urldate = {2021-10-10},
  archiveprefix = {arXiv},
  langid = {english}
}

@article{Chishtie:2012sq,
  title = {Path {{Integral Quantization}} of the {{First Order Einstein-Hilbert Action}} from Its {{Canonical Structure}}},
  author = {Chishtie, Farrukh and McKeon, D.G.C.},
  year = 2012,
  journal = {Classical and Quantum Gravity},
  volume = {29},
  number = {23},
  eprint = {1207.2302},
  primaryclass = {hep-th},
  pages = {235016},
  issn = {0264-9381, 1361-6382},
  doi = {10.1088/0264-9381/29/23/235016},
  urldate = {2021-10-10},
  archiveprefix = {arXiv}
}

@article{Taylor:1971ff,
  title = {Ward {{Identities}} and {{Charge Renormalization}} of the {{Yang-Mills Field}}},
  author = {Taylor, J.C.},
  year = 1971,
  journal = {Nuclear Physics B},
  volume = {33},
  number = {2},
  pages = {436--444},
  issn = {05503213},
  doi = {10.1016/0550-3213(71)90297-5},
  urldate = {2023-12-07},
  copyright = {https://www.elsevier.com/tdm/userlicense/1.0/},
  langid = {english}
}

@incollection{Zinn-Justin:1974ggz,
  title = {Renormalization of {{Gauge Theories}}},
  booktitle = {Lect. {{Notes Phys}}.},
  author = {{Zinn-Justin}, Jean},
  year = 1975,
  volume = {37},
  pages = {1--39},
  publisher = {Springer Berlin Heidelberg},
  address = {Berlin, Heidelberg},
  doi = {10.1007/3-540-07160-1_1},
  urldate = {2023-12-10},
  isbn = {978-3-540-07160-0},
  langid = {english}
}

@article{Slavnov:1972fg,
  title = {Ward {{Identities}} in {{Gauge Theories}}},
  author = {Slavnov, A.A.},
  year = 1972,
  journal = {Theoretical and Mathematical Physics},
  volume = {10},
  pages = {99--107},
  doi = {10.1007/BF01090719}
}

@book{itzyksonQuantumFieldTheory1980,
  title = {Quantum {{Field Theory}}},
  author = {Itzykson, C. and Zuber, J. B.},
  year = 1980,
  series = {International {{Series In Pure}} and {{Applied Physics}}},
  publisher = {McGraw-Hill},
  address = {New York},
  isbn = {978-0-486-44568-7}
}

@article{Brandt:2025lkd,
  title = {Equivalence of First- and Second-Order Formulations of the {{Einstein-Hilbert}} Theory},
  author = {Brandt, F. T. and Frenkel, J. and {Martins-Filho}, S. and McKeon, D. G. C.},
  year = 2026,
  month = jan,
  journal = {Physical Review D},
  volume = {113},
  number = {2},
  eprint = {2510.17615},
  primaryclass = {hep-th},
  pages = {026001},
  issn = {2470-0010, 2470-0029},
  doi = {10.1103/x6h1-bg4n},
  urldate = {2026-03-04},
  archiveprefix = {arXiv},
  langid = {english}
}

@article{Fukushima:2024nxm,
  title = {{{4D Chern-Simons}} Theory with Auxiliary Fields},
  author = {Fukushima, Osamu and Yoshida, Kentaroh},
  year = 2025,
  month = sep,
  journal = {Journal of High Energy Physics},
  volume = {09},
  number = {9},
  eprint = {2407.02204},
  primaryclass = {hep-th},
  pages = {001},
  issn = {1029-8479},
  doi = {10.1007/JHEP09(2025)001},
  urldate = {2026-03-07},
  archiveprefix = {arXiv}
}

@article{Ferko:2024ali,
  title = {Infinite {{Family}} of {{Integrable Sigma Models Using Auxiliary Fields}}},
  author = {Ferko, Christian and Smith, Liam},
  year = 2024,
  month = sep,
  journal = {Physical Review Letters},
  volume = {133},
  number = {13},
  eprint = {2405.05899},
  primaryclass = {hep-th},
  pages = {131602},
  publisher = {American Physical Society},
  doi = {10.1103/PhysRevLett.133.131602},
  urldate = {2026-03-07},
  archiveprefix = {arXiv}
}

@article{Batalin:1983wj,
  title = {{{FEYNMAN RULES FOR REDUCIBLE GAUGE THEORIES}}},
  author = {Batalin, I. A. and Vilkovisky, G. A.},
  year = 1983,
  journal = {Phys. Lett. B},
  volume = {120},
  pages = {166--170},
  doi = {10.1016/0370-2693(83)90645-7}
}

@article{Upadhyay:2013xaa,
  title = {Perturbative Quantum Gravity in {{Batalin-Vilkovisky}} Formalism},
  author = {Upadhyay, Sudhaker},
  year = 2013,
  month = jun,
  journal = {Physics Letters B},
  volume = {723},
  number = {4-5},
  eprint = {1305.4709},
  primaryclass = {hep-th},
  pages = {470--474},
  issn = {03702693},
  doi = {10.1016/j.physletb.2013.05.051},
  urldate = {2026-03-08},
  archiveprefix = {arXiv},
  langid = {english}
}

@article{Cattaneo:2017kkv,
  title = {{{BV-BFV}} Approach to {{General Relativity}}: {{Palatini}}--{{Cartan}}--{{Holst}} Action},
  shorttitle = {{{BV-BFV}} Approach to {{General Relativity}}},
  author = {Cattaneo, A. S. and Schiavina, M.},
  year = 2020,
  month = may,
  journal = {Advances in Theoretical and Mathematical Physics},
  volume = {23},
  number = {8},
  eprint = {1707.06328},
  primaryclass = {math-ph},
  pages = {2025--2059},
  publisher = {International Press of Boston},
  issn = {1095-0753},
  doi = {10.4310/ATMP.2019.v23.n8.a3},
  urldate = {2026-03-08},
  archiveprefix = {arXiv},
  langid = {english}
}

@article{Piguet:2000fy,
  title = {Ghost Equations and Diffeomorphism-Invariant Theories},
  author = {Piguet, Olivier},
  year = 2000,
  journal = {Classical and Quantum Gravity},
  volume = {17},
  number = {18},
  eprint = {hep-th/0005011},
  pages = {3799--3806},
  issn = {0264-9381},
  doi = {10.1088/0264-9381/17/18/314},
  urldate = {2026-03-08},
  archiveprefix = {arXiv},
  langid = {english}
}

@article{Alvarez:2025hym,
  title = {One Loop Analysis of the Cubic Action for Gravity},
  author = {{\'A}lvarez, Enrique and Anero, Jes{\'u}s and Martin, Carmelo P.},
  year = 2025,
  month = apr,
  journal = {Journal of High Energy Physics},
  volume = {04},
  number = {4},
  eprint = {2502.08434},
  primaryclass = {hep-th},
  pages = {117},
  issn = {1029-8479},
  doi = {10.1007/JHEP04(2025)117},
  urldate = {2026-03-08},
  archiveprefix = {arXiv},
  langid = {english}
}

@article{Ashtekar:1986yd,
  title = {New {{Variables}} for {{Classical}} and {{Quantum Gravity}}},
  author = {Ashtekar, A.},
  year = 1986,
  journal = {Physical Review Letters},
  volume = {57},
  number = {18},
  pages = {2244--2247},
  publisher = {American Physical Society},
  doi = {10.1103/PhysRevLett.57.2244},
  urldate = {2026-03-08}
}

@article{Immirzi:1996di,
  title = {Real and Complex Connections for Canonical Gravity},
  author = {Immirzi, Giorgio},
  year = 1997,
  journal = {Classical and Quantum Gravity},
  volume = {14},
  number = {10},
  eprint = {gr-qc/9612030},
  pages = {L177-L181},
  issn = {0264-9381},
  doi = {10.1088/0264-9381/14/10/002},
  urldate = {2026-03-08},
  archiveprefix = {arXiv},
  langid = {english}
}

@article{BarberoG:1994eia,
  title = {Real {{Ashtekar Variables}} for {{Lorentzian Signature Space-times}}},
  author = {Barbero, J. Fernando},
  year = 1995,
  journal = {Physical Review D},
  volume = {51},
  number = {10},
  eprint = {gr-qc/9410014},
  pages = {5507--5510},
  issn = {0556-2821},
  doi = {10.1103/PhysRevD.51.5507},
  urldate = {2026-03-08},
  archiveprefix = {arXiv}
}

@article{Holst:1995pc,
  title = {Barbero's {{Hamiltonian}} Derived from a Generalized {{Hilbert-Palatini}} Action},
  author = {Holst, S{\"o}ren},
  year = 1996,
  journal = {Physical Review D},
  volume = {53},
  number = {10},
  eprint = {gr-qc/9511026},
  pages = {5966--5969},
  publisher = {American Physical Society},
  doi = {10.1103/PhysRevD.53.5966},
  urldate = {2026-03-08},
  archiveprefix = {arXiv}
}

@article{Bergmann:1949zz,
  title = {Non-{{Linear Field Theories}}},
  author = {Bergmann, Peter G.},
  year = 1949,
  journal = {Physical Review},
  volume = {75},
  number = {4},
  pages = {680--685},
  publisher = {American Physical Society},
  doi = {10.1103/PhysRev.75.680},
  urldate = {2026-03-08}
}

@article{Dirac:1950pj,
  title = {Generalized {{Hamiltonian Dynamics}}},
  author = {Dirac, P. A. M.},
  year = 1950,
  journal = {Canadian Journal of Mathematics},
  volume = {2},
  pages = {129--148},
  issn = {0008-414X, 1496-4279},
  doi = {10.4153/CJM-1950-012-1},
  urldate = {2026-03-08},
  langid = {english}
}

@article{stueckelberg1957,
  title = {Theory of the Radiation of Photons of Small Arbitrary Mass},
  author = {Stueckelberg, E. C. G.},
  year = 1957,
  journal = {Helv. Phys. Acta},
  volume = {30},
  pages = {209--215},
  doi = {10.5169/seals-112814}
}

@article{Henneaux:1989jq,
  title = {Lectures on the {{Antifield-BRST Formalism}} for {{Gauge Theories}}},
  author = {Henneaux, Marc},
  year = 1990,
  journal = {Nuclear Physics B - Proceedings Supplements},
  volume = {18},
  number = {1},
  pages = {47--106},
  issn = {09205632},
  doi = {10.1016/0920-5632(90)90647-D},
  urldate = {2026-03-08},
  copyright = {https://www.elsevier.com/tdm/userlicense/1.0/},
  langid = {english}
}

@article{Cattaneo:2015xca,
  title = {{{BV-BFV}} Approach to {{General Relativity}}, {{Einstein-Hilbert}} Action},
  shorttitle = {{{BV-BFV}} Approach to General Relativity},
  author = {Cattaneo, Alberto S. and Schiavina, Michele},
  year = 2016,
  month = feb,
  journal = {Journal of Mathematical Physics},
  volume = {57},
  number = {2},
  eprint = {1509.05762},
  primaryclass = {math-ph},
  pages = {023515},
  issn = {0022-2488, 1089-7658},
  doi = {10.1063/1.4941410},
  urldate = {2026-04-22},
  archiveprefix = {arXiv},
  langid = {english}
}

@article{Bashkirov:2005ig,
  title = {{{ON THE BV QUANTIZATION OF GAUGE GRAVITATION THEORY}}},
  author = {Bashkirov, D. and Sardanashvily, G.},
  year = 2005,
  month = apr,
  journal = {International Journal of Geometric Methods in Modern Physics},
  volume = {02},
  number = {02},
  pages = {203--226},
  issn = {0219-8878, 1793-6977},
  doi = {10.1142/S021988780500051X},
  urldate = {2026-04-22},
  langid = {english}
}

@article{Pottel:2020iuz,
  title = {Perturbative Quantization of {{Einstein-Hilbert}} Gravity Embedded in a Higher Derivative Model},
  author = {Pottel, Steffen and Sibold, Klaus},
  year = 2021,
  month = oct,
  journal = {Physical Review D},
  volume = {104},
  number = {8},
  pages = {086012},
  issn = {2470-0010, 2470-0029},
  doi = {10.1103/PhysRevD.104.086012},
  urldate = {2026-04-22},
  langid = {english}
}

@article{Barnich:1994db,
  title = {Local {{BRST}} Cohomology in the Antifield Formalism: {{I}}. {{General}} Theorems},
  shorttitle = {Local {{BRST}} Cohomology in the Antifield Formalism},
  author = {Barnich, Glenn and Brandt, Friedemann and Henneaux, Marc},
  year = 1995,
  month = nov,
  journal = {Communications in Mathematical Physics},
  volume = {174},
  number = {1},
  pages = {57--91},
  issn = {0010-3616, 1432-0916},
  doi = {10.1007/BF02099464},
  urldate = {2026-04-22},
  copyright = {http://www.springer.com/tdm},
  langid = {english}
}

@article{Brunetti:2013maa,
  title = {Quantum {{Gravity}} from the {{Point}} of {{View}} of {{Locally Covariant Quantum Field Theory}}},
  author = {Brunetti, Romeo and Fredenhagen, Klaus and Rejzner, Katarzyna},
  year = 2016,
  month = aug,
  journal = {Communications in Mathematical Physics},
  volume = {345},
  number = {3},
  pages = {741--779},
  issn = {0010-3616, 1432-0916},
  doi = {10.1007/s00220-016-2676-x},
  urldate = {2026-04-22},
  langid = {english}
}

@article{Gomis:1994he,
  title = {Antibracket, Antifields and Gauge-Theory Quantization},
  author = {Gomis, Joaquim and Par{\'i}s, Jordi and Samuel, Stuart},
  year = 1995,
  month = aug,
  journal = {Physics Reports},
  volume = {259},
  number = {1-2},
  pages = {1--145},
  issn = {03701573},
  doi = {10.1016/0370-1573(94)00112-G},
  urldate = {2026-04-22},
  copyright = {https://www.elsevier.com/tdm/userlicense/1.0/},
  langid = {english}
}

@article{Nojiri:2017ncd,
  title = {Modified Gravity Theories on a Nutshell: {{Inflation}}, Bounce and Late-Time Evolution},
  shorttitle = {Modified Gravity Theories on a Nutshell},
  author = {Nojiri, S. and Odintsov, S.D. and Oikonomou, V.K.},
  year = 2017,
  month = jun,
  journal = {Physics Reports},
  volume = {692},
  pages = {1--104},
  issn = {03701573},
  doi = {10.1016/j.physrep.2017.06.001},
  urldate = {2026-04-22},
  langid = {english}
}

@article{Oikonomou:2025htz,
  title = {Strong Gravity Effects on {{R}} 2 -Corrected Single Scalar Field Inflation and Compatibility with the {{ACT}} Data},
  author = {Oikonomou, V.K.},
  year = 2025,
  month = dec,
  journal = {Physics Letters B},
  volume = {871},
  pages = {139972},
  issn = {03702693},
  doi = {10.1016/j.physletb.2025.139972},
  urldate = {2026-04-22},
  langid = {english}
}

@article{Kupka:2025hln,
  title = {Batalin-{{Vilkovisky Formulation}} of {{N}} = 1 {{Supergravity}} in {{Ten Dimensions}}},
  author = {Kupka, Julian and {Strickland-Constable}, Charles and Valach, Fridrich},
  year = 2025,
  month = may,
  journal = {Physical Review Letters},
  volume = {134},
  number = {21},
  pages = {211602},
  issn = {0031-9007, 1079-7114},
  doi = {10.1103/PhysRevLett.134.211602},
  urldate = {2026-04-22},
  langid = {english}
}

@article{Hu:2023juh,
  title = {The Effective Field Theory Approach to the Strong Coupling Issue in f({{T}}) Gravity},
  author = {Hu, Yu-Min and Zhao, Yaqi and Ren, Xin and Wang, Bo and Saridakis, Emmanuel N. and Cai, Yi-Fu},
  year = 2023,
  month = jul,
  journal = {Journal of Cosmology and Astroparticle Physics},
  volume = {07},
  number = {07},
  eprint = {2302.03545},
  primaryclass = {gr-qc},
  pages = {060},
  issn = {1475-7516},
  doi = {10.1088/1475-7516/2023/07/060},
  urldate = {2026-04-25},
  archiveprefix = {arXiv}
}

@article{Buchdahl:1970ldb,
  title = {Non-{{Linear Lagrangians}} and {{Cosmological Theory}}},
  author = {Buchdahl, H.A.},
  year = 1970,
  month = sep,
  journal = {Monthly Notices of the Royal Astronomical Society},
  volume = {150},
  number = {1},
  pages = {1--8},
  issn = {0035-8711, 1365-2966},
  doi = {10.1093/mnras/150.1.1},
  urldate = {2026-04-25},
  langid = {english}
}

@article{Asorey:2025bel,
  title = {Reflection Positivity in a Higher-Derivative Model with Physical Bound States of Ghosts},
  author = {Asorey, Manuel and Krein, Gast{\~a}o and Pardina, Miguel and Shapiro, Ilya L.},
  year = 2026,
  month = feb,
  journal = {Journal of High Energy Physics},
  volume = {02},
  number = {2},
  eprint = {2511.15283},
  primaryclass = {hep-th},
  pages = {020},
  issn = {1029-8479},
  doi = {10.1007/JHEP02(2026)020},
  urldate = {2026-04-25},
  archiveprefix = {arXiv},
  langid = {english}
}

@article{Batalin:1991jm,
  title = {Existence Theorem for the Effective Gauge Algebra in the Generalized Canonical Formalism with {{Abelian}} Conversion of Second Class Constraints},
  author = {Batalin, I.A. and Tyutin, I.V.},
  year = 1991,
  journal = {International Journal of Modern Physics A},
  volume = {6},
  number = {18},
  pages = {3255--3282},
  issn = {0217-751X, 1793-656X},
  doi = {10.1142/S0217751X91001581},
  urldate = {2026-04-28},
  langid = {english}
}

\end{document}